\documentclass[preprint,showpacs,preprintnumbers,amsmath,amssymb]{revtex4}

\usepackage{graphicx}
\usepackage{dcolumn}
\usepackage{bm,color}
\usepackage{epsfig,subfigure,amsmath}

\begin{document}

\preprint{ArXiv v.2}

\title{Random Lattice Gauge Theories and Differential Forms\footnote{A modified version of this manuscript appears 
in {\it ISRN Mathematical Physics}, vol. 2013, 487270 (2013). doi:10.1155/2013/487270}}


\author{F. L. Teixeira}
\affiliation{ElectroScience Laboratory, Department of Electrical and Computer Engineering
\\The Ohio State University, Columbus Ohio 43212, USA.}%

\date{\today}


\begin{abstract}

We provide a brief overview on the application of the exterior calculus of differential forms to the {\it ab initio} formulation of field theories based upon random simplicial lattices. In this framework, discrete analogues of the exterior derivative and the Hodge star operator are employed for the factorization of discrete field equations into a purely combinatorial (metric-free) part and a metric-dependent part. The Hodge star duality (isomorphism) is invoked to motivate the use of primal and dual lattices (a dual cell complex). The natural role of Whitney forms in the construction of discrete Hodge star operators is stressed.

\end{abstract}



\maketitle


\section{Introduction}

The need to formulate field theories on a lattice (mesh, grid) arises from two main reasons, which may occur simultaneously or not. First, the lattice provides a natural `regularization' of divergences in lieu of renormalization techniques~\cite{Montvay}. Such regularization does not need to be viewed as an {\it ad hoc} step, but instead as a natural consequence of assuming the field theory to be, at some fundamental level, an {\it effective} (`low'-energy) description~\cite{Zee}.  Second, the lattice provides a direct route to compute, in a non-perturbative fashion, quantities of interest by numerical simulations. Nontrivial domains and complex boundary conditions can then be easily treated as well~\cite{Chew},\cite{sanmartin},\cite{Bretones},\cite{TeixeiraAP08}.
For these, the use of irregular (`random') lattices are often of interest to gain geometrical flexibility.
Irregular lattices are also of interest as a means to provide a potentially faster convergence to the continuum limit, near-isotropic lattice dispersion properties, and better `conservation' of some (e.g., long-range translational and rotational) symmetries~\cite{Christ},\cite{Bolander}. In some cases, 
irregular lattices are useful for universality tests as well~\cite{Drouffe},\cite{Gockeler}.

Lattice theories are typically developed by taking the counterpart continuum theory as starting point and then applying discretization 
techniques whereby derivatives are approximated by finite-differences or some constraints
are enforced on the functional space of admissible solutions  to be spanned by a 
{\it finite} set of `basis' functions (e.g., `Galerkin methods' such as spectral elements and finite elements). These discretization strategies have proved very useful in many settings; however, 
they often produce difficulties in the case of irregular (`random') lattices. Among such difficulties are  
($i$) numerical instabilities in marching-on-time algorithms (regardless of the time integration method used), ($ii$) convergence problems in algorithms relying on iterative linear solvers, and ($iii$) spurious (`ghost') modes and/or extraneous degrees of freedom. These problems often 
(but not always) appear associated with    
highly skewed or obtuse lattice elements, or at the boundary between
 heterogeneous (hybrid) lattices subcomponent, comprising overlapped domains 
or ``mesh-stitching'' interfaces, for example. Clearly, such difficulties put a constraint on the geometric flexibility that irregular lattices are intended for, and may require 
stringent (and computationally demanding) mesh quality controls.
These difficulties also impact the ability to utilize `mesh refinement' 
strategies based on a priori error estimates.
The reasons behind these difficulties can be traced to 
an inconsistent rendering of the differential calculus and degrees of freedom  
on the lattice. A rough classification of those inconsistencies
is provided in Appendix D.

The objective of this work is to provide a brief overview on the application of exterior calculus of differential forms 
to the {\it ab initio} formulation of gauge field theories on irregular simplicial (or `random') lattices~\cite{KomorowskiBAPS75},\cite{DodziukAJM76},\cite{SorkinJMP75},\cite{weingarten77},\cite{MullerAM78},\cite{BecherZPC82},\cite{Rabin},\cite{Bossavit98},\cite{BossavitEJM91},\cite{AdamsArXiv}, \cite{MattiussiJCP97},\cite{KettunenMAGN98},\cite{MineP16},\cite{KettunenMAGN99},\cite{SenAdamsPRE00},\cite{ShapiroMCS00},\cite{KotiugaPIER01},\cite{TeixeiraPIER01},\cite{KettunenPIER01},\cite{MineP31},\cite{Wise}. 
In the exterior calculus framework,
the lattice is treated as a cell complex (in the parlance of algebraic topology~\cite{Schwarz94}) instead of simply a collection of discrete points, and 
dynamic fields are represented by means of discrete differential forms (cochains) of various degrees~\cite{KettunenPIER01},\cite{MineP25},\cite{MineP27},\cite{Wise}.
This prescription provides a basis for developing a consistent `discrete calculus' on 
irregular lattices, and discrete analogues to partial differential equations
that better adheres to the underlying physics.

This topic intersects many disparate application areas. For concreteness, we use 
classical electrodynamics in 3+1 dimensions as a basic model. Although some familiarity with the exterior calculus of differential forms is assumed~\cite{Bossavit98},\cite{BossavitEJM91},\cite{Whitney57},\cite{Misner},\cite{DeschampsIEEE82},\cite{KotiugaJAP93},\cite{MineP15},\cite{MineP17},\cite{MineP29}, the discussion is mostly kept at a tutorial level. Finally, we stress that this is a review paper and no claim of originality is intended.

\section{Pre-metric lattice equations}

Let us denote the space of differential $p$-forms on a smooth connected manifold 
$\Omega$ as $\Lambda^p(\Omega)$.
From a geometric perspective, a differential $p$-form $\alpha^p \in \Lambda^p(\Omega)$ can be viewed as an {\it oriented $p$-dimensional density}, or an object naturally associated with $p$-dimensional domains of integration $U_p$ such that the lattice contraction (`pairing') below:
\begin{equation}
\left<U_p \, ,\alpha^p\right> \doteq \int_{U_p} \alpha^p
\label{pairing}
\end{equation}
gives a real number (in our context) for each choice of $U_p$~\cite{MineP16}. On a lattice $\mathcal{K}$, $U_p$ is restricted to be a union of elements from the {\it finite} set of $p$-dimensional $N_p$ oriented lattice elements, which we denote $\Gamma_p(\mathcal{K}) = \{\sigma_{p,i} \, , i=1,\ldots,N_p\}$. These are 
collective called `$p$-chains'. In four-dimensions for example, they correspond to the possible 
unions of elements from the set of vertices (nodes) $\sigma_0$, edges (`links') $\sigma_1$, facets (`plaquettes') $\sigma_2$, volume cells (`voxels') $\sigma_3$, and hypervolume cells $\sigma_4$, for $p=1,\ldots,4$, respectively.  In the discrete setting, the degrees of freedom are reduced to the set of pairings (\ref{pairing}) on each one of the lattice elements. 

On the lattice, the pairing above can be understood as a map 
${\mathcal R}^p: \Lambda^p(\Omega) \rightarrow \Gamma^p(\mathcal{K})$ such that
\begin{equation}
{\mathcal R}^p (\alpha_{p}) =  \left<\sigma_{p,i} \, ,\alpha^p\right> \doteq \int_{\sigma_{p,i}} \alpha^p
\label{pairing2}
\end{equation}
defines its action on the basis of $p$-chains. Note that we use $\Gamma^p(\mathcal{K})$ to denote the space dual to $\Gamma_p(\mathcal{K})$, i.e. the space $p$-{\it cochains}. The latter can be viewed as the space of `discrete differential forms'. Because of this, and with some abuse of language, we 
use the terminology `differential forms' and `cochains' interchangeably 
to denote the same objects in what follows. The map
${\mathcal R}^p$ is called the {\it de Rham map}~\cite{MineP16}.

The basic differential operator of exterior calculus is the exterior derivative $d$, applicable to any number of dimensions. The discretization of $d$ on a general irregular lattice can be effected by a
straightforward application of the generalized Stokes' theorem~\cite{MineP16}
\begin{equation}
\int_{\sigma_{p+1}} d \, \alpha^p = \int_{\partial \sigma_{p+1}} \alpha^p
\label{stokes}
\end{equation}
with $p=0, \ldots, 3$ in $n=4$. 
In the above, $\partial$ is the boundary operator, which simply maps a $p$-dimensional lattice element to the set of $(p-1)$-dimensional lattice elements that comprise its boundary, preserving orientation. This theorem sets $\partial$ as the formal adjoint of $d$ in terms of the pairing given in (\ref{pairing}), that is 
$\left< \sigma_{p+1}, d \alpha^p \right> = \left< \partial \sigma_{p+1}, \alpha^p \right>$. 
Computationally, the boundary operator can be implemented by 
means of incidence matrices~\cite{MineP16},\cite{MineP31},\cite{Guth} such that
\begin{equation}
\partial \, \sigma_{p+1,i} = \sum_j C_{ij}^{p} \, \sigma_{p,j}
\label{bofb}
\end{equation}
where the indices $i$ and $j$ run over all $(p+1)$- and $p$-dimensional lattice elements, respectively. The incidence matrix entries are such that $C_{ij}^p \in \{-1,0,1\}$ for all $p$, with sign determined by the relative orientation of lattice elements $i$ and $j$. The restriction to this set of integer values reflects the `metric-free' nature of the exterior derivative: only information about element {\it connectivity}, that is, the combinatorial aspects of the lattice, is involved here. It turns out that the metric is fully encoded by Hodge star operators, the discretization of which will be discussed further down below.

Using eqs.~(\ref{stokes}) and (\ref{bofb}), one can write
\begin{equation}
\int_{\sigma_{p+1,i}} d \, \alpha^p = \sum_j C_{ij}^p \int_{\sigma_{p,j}} \alpha^p
\end{equation}
for all $i$,
so that the derivative operation is replaced by a proper sum over $j$.  On the lattice, the nilpotency of the operators ${\partial} \circ {\partial}=d \circ d=0$ \cite{Kheyfets} is recovered by the constraint~\cite{MineP16}
\begin{equation}
\sum_k C_{ik}^{p+1} \, C_{kj}^p = 0
\end{equation}
for all $i$ and $j$.

\section{Example: Lattice electrodynamics}

We write Maxwell's equations in a four-dimensional Lorentzian manifold $\Omega$ as~\cite{Misner}
\begin{equation}
d F=0 
\end{equation}
\begin{equation}
d G=*{\mathcal J}
\end{equation}
where $d$ is the four-dimensional exterior derivative, $F$ and $G$ are the so-called Faraday and Maxwell 2-forms, respectively, and $*{\mathcal J}$ is the charge-current density 3-form. The Hodge star operator $*$ is an isomorphism that maps $p$-forms to $(4-p)$-forms, and more generally $p$ forms to $(n-p)$ forms in a $n$-dimensional manifold, and, as mentioned before, depends on the metric of $\Omega$~\cite{MineP16},\cite{KettunenMAGN99},\cite{Misner},\cite{DeschampsIEEE82},\cite{HiptmairNM01},\cite{MineP26},\cite{MineP28},\cite{MineP30}. The above equations are complemented by the relation $G=*F$, which indicates that 
$F$ and $G$ are `Hodge duals' of each other.

\subsection{Primal and dual lattices}

Since $F$ and $G$ are 2-forms, they should be discretized as 2-cochains residing on plaquettes (2-chains) of the 4-dimensional lattice; however, it is important to recognize that these two forms are of different types: $F$ is a `ordinary' (or
`non-twisted') differential form, whereas $G$ (as well as $*{\mathcal J}$) is a `twisted' (or `odd') 
differential form~\cite{burke}. The basic difference here has to do with orientation: ordinary forms have internal orientation whereas twisted forms have external orientation~\cite{MattiussiJCP97},\cite{MineP16},\cite{burke},\cite{tonti76},\cite{TontiPIER01}. These two types of orientations exhibit different symmetries under reflection, a distinction akin to that between proper (or polar) tensors and pseudo (or axial) tensors. 
Only twisted forms admit integration in non-orientable manifolds.
These two types of forms are associated with two distinct `cell complexes' (lattices), each one inheriting the corresponding orientation: the ordinary form $F$ is associated with the set of plaquettes $\Gamma_2$ on the  `ordinary cell complex' $\mathcal{K}$, thus belonging to $ 
\Gamma^{2}(\mathcal{K})$, while the twisted forms $G$ and $*{\mathcal J}$ are associated with the set of plaquettes ${\tilde \Gamma}_2$ on the `twisted cell complex' $\tilde{\mathcal{K}}$~\cite{MineP16},\cite{TeixeiraPIER01},\cite{TontiPIER01},\cite{TontiRAL72}, thus belonging to
 $\Gamma^{2}(\tilde{\mathcal{K}})$. Consequently, we also have two sets of
incidence matrices $C_{ij}^p$ and $\tilde{C}_{ij}^p$, one for each lattice. 
It is convenient to denote $\mathcal{K}$ as the `primal lattice'  and $\tilde{\mathcal{K}}$ as the `dual lattice'~\cite{MineP16}. 

As detailed further below, these two lattices become intertwined by the Hodge duality $F=*G$. 
 The need for dual lattices can be motivated from a combinatorial standpoint~\cite{AdamsArXiv},\cite{SenAdamsPRE00} or from a computational standpoint (to provide higher-order convergence to the continuum, for example)~\cite{YeeAP69},\cite{Taflove95},\cite{NicolaidesSIAM97}. 
The importance of a primal/dual lattice setup for the discretization of the Hodge star operator in the context of field theories was first recognized in~\cite{AdamsArXiv}, 
where it was shown that such setup is also crucial for correctly reproducing topological invariants in the discrete setting. 

\subsection{3+1 theory}

At this point, it is suitable to degeometrize time and treat it simply as a parameter. This corresponds to the majority of low-energy applications involving Maxwell's equations, in which one is interested
in predicting the field evolution along different spatial slices for a given set of initial and boundary 
conditions.
In this case, we still use the symbols $\mathcal{K}$ and $\tilde{\mathcal{K}}$ for the primal and dual lattices, but they now refer to three-dimensional {\it spatial} lattices.
Similarly, $\Omega$ now refers to a three-dimensional Euclidean manifold .
In such a 3+1 setting, one can decompose $F$ and $G$ as
 \begin{equation}
 F = E \wedge dt + B
 \end{equation}
  \begin{equation}
 G = D - H \wedge dt
 \end{equation}
and the source density as
  \begin{equation}
 *{\mathcal J} = -J \wedge dt  + \rho
 \end{equation}
 where $\wedge$ is the wedge product,
 $E$ and $H$ are the electric intensity and magnetic intensity 1-forms 
on $\Gamma_1$ and
${\tilde \Gamma}_1$ respectively, $D$ and $B$ are the electric flux and magnetic flux 2-forms
on
${\tilde \Gamma}_2$ and $\Gamma_2$ respectively, 
$J$ is the electric current density 2-form on
${\tilde \Gamma_2}$
, and $\rho$ is the electric charge density 3-form on ${\tilde \Gamma_3}$ (corresponding assignments for the 2+1 and 1+1 cases are provided in~\cite{MineP27}). 
As a result, Maxwell's equations reduce to
 \begin{equation}
 dE=-\partial_t B
 \label{faraday}
 \end{equation}  
\begin{equation} 
 dH=\partial_t D +J
\label{ampere}
\end{equation}
 representing Faraday's and Ampere's law, respectively.
Here, $d$ stands for the 3-dimensional spatial exterior derivative. Note that both eqs. (\ref{faraday}) and (\ref{ampere}) are metric-free.
 They are supplemented by Hodge star relations given by
  \begin{equation}
  D = \star_\epsilon E
  \end{equation}  
\begin{equation}
  H =\star_{\mu^{-1}} B
  \end{equation}
  now involving two Hodge star maps in three-dimensional space: 
$\star_\epsilon: \Lambda^1(\Omega) \rightarrow \Lambda^{2}(\Omega)$
and  $\star_{\mu^{-1}}: \Lambda^2(\Omega) \rightarrow \Lambda^{1}(\Omega)$. On the lattice, we have the corresponding discrete counterparts:
$[\star_\epsilon]: \Gamma^1(\mathcal{K}) \rightarrow \Gamma^{2}(\tilde{\mathcal{K}})$
and  $[\star_{\mu^{-1}}]: \Gamma^2(\mathcal{K}) \rightarrow \Gamma^{1}(\tilde{\mathcal{K}})$.
The subscripts $\epsilon$ and $\mu$ in $\star_\epsilon$ and $\star_{\mu^{-1}}$ serve to indicate that these operators also incorporate {\it macroscopic} constitutive material properties 
through the local permittivity and permeability values~\cite{codecasa07} (we assume dispersionless media for simplicity). In Riemannian manifolds (and in particular, Euclidean space) and reciprocal media, these two Hodge star operators are symmetric and positive-definite~\cite{Auchmann}.

In what follows, we employ the following short-hand notation for cochains:
$\left<\sigma_{1,i},E\right>=E_i$, 
$\left<{\tilde \sigma}_{1,i},H\right>=H_i$,
$\left<\tilde{\sigma}_{2,i},D\right>=D_i$,
$\left<\sigma_{2,i},B\right>=B_i$,
 $\left<\tilde{\sigma}_{2,i},J\right>=J_i$, and
 $\left<\tilde{\sigma}_{3,i},\rho\right>=\rho_i$, where the indices run over the respective basis of $p$-chains in either $\mathcal K$ or ${\tilde{\mathcal K}}$, $p=1,2,3$. With the exception of Appendix A, we restrict ourselves to the 3+1 setting throughout the remainder of this paper. 

\section{Casting the metric on a lattice}

\subsection{Whitney forms}

The Whitney map ${\mathcal W}: \Gamma^p(\mathcal{K}) \rightarrow \Lambda^p(\Omega)$ is the right-inverse of the de Rham map (\ref{pairing2}), that is,
${\mathcal R} \circ {\mathcal W} = \mathcal{I}$, where $\mathcal{I}$ is the identity operator. 
In simplicial lattices, this morphism can be constructed using the so-called Whitney forms~\cite{MullerAM78},\cite{MineP16},\cite{KotiugaJAP93},\cite{MineP26},\cite{BossavitIEE88},\cite{kotiugabook},\cite{bossavit05},\cite{bohethesis},\cite{HiptmairPIER01},\cite{rapetti},\cite{kangas07} which are basic interpolants from cochains to differential forms~\cite{Whitney57} (other interpolants are also possible~\cite{Buffa11},\cite{Back12}).
By definition, all cell elements of a simplicial lattice are {\it simplices}, i.e., cells whose boundaries
are the union of a minimal number of lower-dimensional cells. In other words, $0$-simplices are nodes, 1-simplices are links, 2-simplices are triangles, 
3-simplices are tetrahedra, and so on. Note that if the primal lattice is simplicial, the dual lattice is not~\cite{MineP25}.
For a $p$-simplex $\sigma_{p,i}$, the (lowest-order) Whitney form is given by 
\begin{equation}
\omega^p [\sigma_{p,i}] \doteq p! \sum_{j=0}^{p} (-1)^i \lambda_{i,j} d\lambda_{i,0} \wedge d\lambda_{i,1} \cdots
d\lambda_{i,j-1} \wedge d\lambda_{i,j+1} \cdots d\lambda_{i,p}
\label{Whitney}
\end{equation}
where $\lambda_{i,j}$, $j=0,\ldots,p$, are the barycentric coordinates associated to $\sigma_{p,i}$.
In the case of a $0$-simplex (node), (\ref{Whitney}) reduces to $\omega^0 [\sigma_{0,i}] = \lambda_{i}$.

From its definition, it is clear that Whitney forms have compact support. Among its important structural properties are:
\begin{equation}
\left<\sigma_{p,i},\omega^p [\sigma_{p,j}] \right> = \int_{\sigma_{p,i}} \omega^p [\sigma_{p,j}] = \delta_{ij}
\label{struc1}
\end{equation}
where $\delta_{ij}$ is the Kronecker delta, which is simply a restatement of ${\mathcal R} \circ {\mathcal W} = \mathcal{I}$, and
\begin{equation}
\omega^{p} [\partial^T \sigma_{p-1,i}] = d\left( \omega^{p-1} [\sigma_{p-1,i}] \right)
\label{struc2}
\end{equation}
where $\partial^T$ is the coboundary operator~\cite{kotiugabook}, consistent with the generalized Stokes' theorem. Further structural properties are provided in~\cite{bossavit05},\cite{bohethesis}.
 Higher-order version of Whitney forms also exist~\cite{HiptmairPIER01},\cite{rapetti}. The key result ${\mathcal W} \circ {\mathcal R} \rightarrow \mathcal{I}$ holds in the limit of zero lattice spacing. This is discussed, together with other related convergence results in various contexts, in~\cite{MullerAM78},\cite{Whitney57},\cite{albeverio90},\cite{albeverio95},\cite{wilson07},\cite{wilson11},\cite{halvorsen12}.

Using the short-hand $\omega^p [\sigma_{p,i}] = \omega^p_i$, we can write the following expansions for $E$ and $B$ in a irregular simplicial lattice, in terms of its cochain representations:
\begin{equation}
E = \sum_i E_i \, \omega^1_i
\label{exp1}
\end{equation}
\begin{equation}
B = \sum_i B_i \, \omega^2_i
\label{exp2}
\end{equation}
where the sums run over all primal lattice edges and faces, respectively. 

One could argue that Whitney forms are continuum objects that should have no fundamental 
place on a truly discrete theory. In our view, this is only partially true. In many applications (see, for example, the discussion on space-charge effects below), it is less natural to consider the lattice as endowed with some a priori discrete metric structure than it is to consider it instead as embedded in an underlying continuum (say, Euclidean) manifold with metric and hence inheriting all metric properties from it. In the latter case, Whitney forms provide the standard route to incorporate metric information into the discrete Hodge star operators, as described next. 

\subsection{Discrete Hodge star operator}

In a source-free media, we can write the Hamiltonian as
\begin{equation}
{\mathcal H}=\frac{1}{2} \int_{\Omega} \left( E \wedge D + H \wedge B \right)=
\int_{\Omega} \left( E \wedge \star_{\epsilon} E + \star_{\mu^{-1}} B \wedge B \right)
\end{equation}
Using eqs.~(\ref{exp1}) and (\ref{exp2}), the lattice Hamiltonian assumes the expected quadratic form:
\begin{equation}
{\mathcal H}=
\sum_i \sum_j E_i \, [\star_{\epsilon}]_{ij} \, E_j + 
\sum_i \sum_j B_i \, [\star_{\mu^{-1}}]_{ij} \, B_j
\end{equation}
where we immediately identify the symmetric positive definite matrices
\begin{equation}
[\star_{\epsilon}]_{ij} =  \int_{\Omega} \omega^1_i \wedge \star_{\epsilon} \omega^1_j 
\label{hodge1m}
\end{equation}
\begin{equation}
[\star_{\mu^{-1}}]_{ij} = \int_{\Omega} \left(
\star_{\mu^{-1}} \omega^2_i \right) \wedge \omega^2_j
\label{hodge2m}
\end{equation}
as the discrete realization of the Hodge star operator(s) on a simplicial lattice~\cite{KettunenMAGN99},\cite{bossavitjapan} so that
\begin{equation}
D_i = \sum_j [\star_{\epsilon}]_{ij} E_j 
\label{hodge1}
\end{equation}
\begin{equation}
H_i = \sum_j [\star_{\mu^{-1}}]_{ij} B_j. 
\label{hodge2}
\end{equation}
From the above, the Hamiltonian can be also expressed as
\begin{equation}
{\mathcal H}=
\sum_i E_i \, D_i + 
\sum_i H_i \, B_i
\end{equation}

\subsection{Symplectic structure and dynamic degrees of freedom}

The Hodge star matrices $[\star_\epsilon]$ and 
$[\star_{\mu^{-1}}]$ have different sizes. 
The number of elements in $[\star_\epsilon]$ is equal to $N_1 \times N_1$, whereas 
the number of elements in $[\star_\mu^{-1}]$ is equal to $N_2 \times N_2$.
In other words,
$\Theta(E) = \Theta(D)
\neq \Theta(B) = \Theta(H)$, where $\Theta$ denotes the number of (discrete) degrees of freedom in the corresponding field.

One important property of a Hamiltonian system is its symplectic character, associated with area
preservation in phase space.  The symplectic character of
the Hamiltonian in principle would require a canonical pair
such as $E,B$ to have identical number of degrees of freedom.
This apparent contradiction can be explained by the fact that Maxwell's equations (\ref{faraday}) and (\ref{ampere}) can be thought as a\textit{\ constrained} dynamic
system (by the divergence conditions) so that, even though $\Theta(E) 
\neq \Theta(B)$, we still have $\Theta^d(E) = \Theta^d(B)$, where $\Theta^d$ denotes the number of {\it dynamic} degrees of freedom. This is discussed further below in Section VI, in connection with the discrete Hodge decomposition on a lattice.

\section{Semi-discrete equations}

\subsection{Local and ultra-local lattice coupling}

By using a contraction in the form of~(\ref{pairing2}) on both sides of~(\ref{faraday}) with every face
$\sigma_{2,j}$ of $\mathcal K$, 
and using the fact that $\left< \sigma_{2,j} , \omega^2_i \right>=\left< \sigma_{1,j}, \omega^1_i \right> = \delta_{ij}$ from (\ref{struc1}),
we get
 \begin{equation}
\left< \sigma_{2,j}, \partial_t B \right>=
 \partial_t \sum_i B_i \left< \sigma_{2,j} , \omega^2_i \right>
 = \partial_t B_j
\end{equation}
and
\begin{equation} 
\left< \sigma_{2,j},dE \right>=\left< \partial \sigma_{2,j}, E \right>=
\sum_i E_i \sum_k C^1_{jk} \left< \sigma_{1,k}, \omega^1_i \right>=
\sum_i C^1_{ji} \, E_i
\end{equation}
so that
 \begin{equation}
- \partial_t B_i = \sum_j C^1_{ij} \, E_j
\label{far1}
 \end{equation}
 where the index $i$ runs over all faces of the primal lattice.
On the dual lattice ${\tilde {\mathcal K}}$, we can similarly contract both sides of eq.~(\ref{ampere}) 
with every dual face ${\tilde \sigma}_{2,j}$ to get
\begin{equation}
\partial_t D_i = \sum_j {\tilde C}^1_{ij} \, H_j
\label{ampx}
 \end{equation}
 where now the index $i$ runs over all faces of the dual lattice. Using 
eqs.~(\ref{hodge1}) and (\ref{hodge2}) and the fact that, in three-dimensions 
${\tilde C}^1_{ij} = C^1_{ji}$ ~\cite{MineP16} (up to possible boundary terms ignored here), we can write the last equation in terms of primal lattice quantities as
  \begin{equation}
\partial_t \sum_j  [\star_{\epsilon}]_{ij} \, E_j  =
 \sum_j C^1_{ji}  \sum_k [\star_{\mu^{-1}}]_{jk} B_k 
 \label{primallat}
 \end{equation}
or, by using the inverse Hodge star matrix $[\star_{\epsilon}]^{-1}_{ij}$, as
   \begin{equation}
\partial_t  E_i  =
 \sum_j \Upsilon_{ij} B_j 
 \label{far2}
 \end{equation}
with
    \begin{equation}
\Upsilon_{ij} \doteq \sum_k \sum_l [\star_{\epsilon}]^{-1}_{ik} \,  C^1_{lk} \, [\star_{\mu^{-1}}]_{lj}
\label{stiffness}
  \end{equation}
  The matrix $[\Upsilon]$ can be viewed as the discrete realization, for $p=2$, of the 
  {\it codifferential} operator $\delta=(-1)^p *^{-1} d \, *$ that maps $p$-forms to $(n-p)$-forms~\cite{DeschampsIEEE82}.

Since the continuum operators $\star_\epsilon$
and  $\star_{\mu^{-1}}$ are local~\cite{burke} and, as seen, Whitney forms (\ref{Whitney})
have local support, it follows that the matrices $[\star_\epsilon]$
and  $[\star_{\mu^{-1}}]$ are sparse, indicative of an {\it ultra-local} coupling (in the terminology of~\cite{Katz1998}). In contrast, the {\it numerical} inverse $[\star_{\epsilon}]^{-1}$ used in eq.~(\ref{stiffness}) is, in general, not sparse so that the field coupling between distant elements is nonzero. The lack of sparsity is a potential bottleneck in practical simulations.
However, because the coupling strength in this case decays exponentially~\cite{MineP31},\cite{MineP28}, we can still say (using again the terminology of~\cite{Katz1998}) that the resulting discrete operator encoded by the matrix in~(\ref{stiffness})
 is local. In practical terms, the exponential decay allows one to set a cutoff on the nonzero elements of $[\star_\epsilon]$, based on element magnitudes or on the sparsity pattern of the original matrix $[\star_\epsilon]$, to build a sparse approximate inverse for $[\star_{\epsilon}]$ and hence recover back an ultra-local representation for $\star_{\epsilon}^{-1}$~\cite{MineP31},\cite{MineP27a}. The sparsity pattern of $[\star_\epsilon]$ encodes the nearest-neighbor edge information of the mesh and, consequently, the sparsity pattern of $[\star_\epsilon]^k$ likewise encodes successive `$k$-level' neighbors. The latter sparsity patterns can be used to build, quite efficiently, sparse approximations for
$[\star_\epsilon]^{-1}$, as detailed in~\cite{MineP31}. Once such sparse representations are obtained, eqs.~(\ref{far1}) and (\ref{far2}) can be used in tandem to construct a marching-on-time algorithm (see Appendix E (a), for example) with a sparse structure and hence amenable for large-scale problems.

\subsection{Barycentric dual and barycentric decomposition lattices}

An alternative approach, aimed at constructing a sparse discrete Hodge star for $\star_{\epsilon}^{-1}$ directly from the dual lattice geometry is described in~\cite{TeixeiraPIER01}, based on earlier ideas exposed in~\cite{AdamsArXiv},\cite{SenAdamsPRE00},\cite{AdamsPRL97}. This approach is based on the fact that both primal $\mathcal{K}$  and dual $\tilde{\mathcal{K}}$ lattices can be decomposed into a third (underlying) lattice $\widehat{\mathcal{K}}$ by means of a barycentric decomposition, see~\cite{SenAdamsPRE00}. The dual lattice $\tilde{\mathcal{K}}$ in this case is called the {\it barycentric dual lattice}~\cite{TeixeiraPIER01},\cite{AdamsPRL97} and the underlying lattice $\widehat{\mathcal{K}}$ is called the {\it barycentric decomposition lattice}. Importantly, $\widehat{\mathcal{K}}$  is simplicial and hence admits Whitney forms built on it using~(\ref{Whitney}). Whitney forms on $\widehat{\mathcal{K}}$ can be used as building blocks to construct (dual) Whitney forms on the (non-simplicial) $\tilde{\mathcal{K}}$, and from that, a sparse inverse discrete Hodge star  $[\star_\epsilon^{-1}]$ using integrals akin to ~(\ref{hodge1m}) and~(\ref{hodge2m}). 
An explicit derivation of such dual lattice Whitney forms is provided in~\cite{buffa}. Furthermore, a recent comprehensive survey of this and other approaches based on dual lattices to construct discrete sparse inverse Hodge stars is provided in~\cite{GilletteCAD11}. A comparison between the properties of a barycentric dual and a circumcentric dual is considered in~\cite{Calcagni}, where it is verified that the former induces a (discrete) Laplacian with better properties (in particular, positivity). 

The barycentric dual lattice has the important property below associated with Whitney forms: 
\begin{equation}
\left<\tilde{\sigma}_{(n-p),i},\star \omega^p [\sigma_{p,j}] \right> = \int_{\tilde{\sigma}_{(n-p),i}} \star \omega^p [\sigma_{p,j}] = \delta_{ij}
\label{struc3}
\end{equation}
where $\star$ stands for the spatial Hodge star operator (distilled from constitutive material properties), and $\tilde{\sigma}_{(n-p),i}$ is the dual element to $\sigma_{p,i}$ on the barycentric dual lattice.
The operator $\star$ is such that
\begin{equation}
\int_{\Omega} \omega^p \wedge \star \omega^p  = \int_{\Omega} |\omega|^2 dv
\label{wstar}
\end{equation}
where $|\omega|^2$ is the two-norm of $\omega^p$ and $dv$ is the volume element. 

The identity (\ref{struc3}) plays the role of structural property (\ref{struc1}), on the dual lattice side. 
We stress that identity (\ref{struc3}) is a distinctively characteristic feature of the barycentric dual lattice not shared by other geometrical constructions for the dual lattice. In other words, compatibility with Whitney forms via (\ref{struc3}) naturally forces one to choose the dual lattice to be the barycentric dual. 

From the above, one can also define a (Hodge) duality operator directly on the space of chains, that is
$\star_K: \Gamma_p(\mathcal{K}) \mapsto \Gamma_{n-p}(\tilde{\mathcal{K}})$ with
$\star_K(\sigma_{p,i})=\tilde{\sigma}_{(n-p),i}$ and
$\star_{\tilde K}: \Gamma_p(\tilde{\mathcal{K}}) \mapsto \Gamma_{n-p}(\mathcal{K})$
with
$\star_K(\tilde{\sigma}_{p,i})=\tilde{\sigma}_{(n-p),i}$, so that $\star_K \star_{\tilde K} = \star_{\tilde K} \star_K =1$. This construction is detailed in~\cite{SenAdamsPRE00}.

\subsection{Galerkin duality}

Even though we have chosen to assign $E$ and $B$ to the primal (simplicial) lattice, and consequently $D$, $H$, $J$, and $\rho$ to the dual (non-simplicial) lattice, the reverse is equally possible. In this case, the fields $D$, $H$ become associated to a simplicial lattice and hence can be expressed in terms of Whitney forms; the expressions dual to (\ref{exp1}) and (\ref{exp2}) are now
\begin{equation}
H = \sum_i H_i \, \omega^1_i
\label{exp1a}
\end{equation}
\begin{equation}
D = \sum_i D_i \, \omega^2_i
\label{exp2a}
\end{equation}
with sums running over primal edges and primal faces, respectively,
and where
\begin{equation}
E_i = \sum_j [\star_{\epsilon^{-1}}]_{ij} D_j 
\label{hodge1d}
\end{equation}
\begin{equation}
B_i = \sum_j [\star_{\mu}]_{ij} H_j 
\label{hodge2d}
\end{equation}
with
\begin{equation}
[\star_{\epsilon}^{-1}]_{ij} =  \int_{\Omega} \left( \star_{\epsilon^{-1}} \omega^2_i \right) \wedge  \omega^2_j 
\label{hodge1a}
\end{equation}
\begin{equation}
[\star_{\mu}]_{ij} = \int_{\Omega}
 \omega^1_i \wedge \star_{\mu} \omega^1_j
\label{hodge2a}
\end{equation}
and the two Hodge star maps now used are such that, in the continuum,
$\star_\epsilon^{-1}: \Lambda^2(\Omega) \rightarrow \Lambda^{1}(\Omega)$
and  $\star_{\mu}: \Lambda^1(\Omega) \rightarrow \Lambda^{2}(\Omega)$, and, on the lattice,
$[\star_\epsilon^{-1}]: \Gamma^2(\mathcal{K}) \rightarrow \Gamma^{1}(\tilde{\mathcal{K}})$
and  $[\star_{\mu}]: \Gamma^1(\mathcal{K}) \rightarrow \Gamma^{2}(\tilde{\mathcal{K}})$.
This alternate choice entails a duality between these two formulations, dubbed `Galerkin duality'. This is explored in more detail in~\cite{MineP28}.

\section{Discrete Hodge decomposition and Euler's formula}

For any $p$-form $\alpha ^{p}$, we can write
\begin{equation}
\alpha ^{p}=d\zeta ^{p-1}+\delta \beta ^{p+1}+\chi ^{p},
\label{generalHelmholtz}
\end{equation}
where $\chi ^{p}$ is a harmonic form~\cite{MineP25}. This {\it Hodge decomposition} is unique.
In the particular case of the
$1$-form $E$, we have
\begin{equation}
E=d\phi +\delta A+\chi,  \label{Helmholtz42}
\end{equation}%
where $\phi $ is a $0$-form and $A$ is a $2$-form, with $d\phi $ representing the static field, $\delta A$
the dynamic field, and $\chi$ the harmonic
field component (if any). In a contractible domain, $\chi $ is identically zero and
the Hodge decomposition simplifies to
\begin{equation}
E=d\phi +\delta A.  \label{Hodge}
\end{equation}%
more usually known as Helmholtz decomposition in three-dimensions.

In the discrete setting, the degrees of freedom of
 $\phi $ are associated to the nodes of the primal lattice. Likewise,
 the degrees of freedom of $A$ are associated to the facets of the primal lattice.
Consequently, we have from (\ref{Hodge}) that
\begin{eqnarray}
\Theta^d\left( E\right)  &=&N_{E}^{h}-N_{V}^{h}  \notag \\
&=&\left( N_{E}^{{}}-N_{E}^{b}\right) -\left(
N_{V}^{{}}-N_{V}^{b}\right)
\notag \\
&=&N_{E}^{{}}-N_{V},  \label{Efreedom}
\end{eqnarray}
where $N_{V}$ is the number of primal nodes, $N_{E}$ the number
of
primal edges, and $N_{F}$ the number of primal facets, with
superscript $b$ standing for boundary (fixed) elements and $h$ for interior (free) elements.

On the other hand, once we identify the lattice as a network of (in general) polyhedra, we can apply Euler's polyhedron
formula on the primal lattice to obtain~\cite{MineP28}
\begin{equation}
N_{V}^{{}}-N_{E}^{{}}=1-N_{F}^{{}}+N_{P}^{{}},  \label{Euler31}
\end{equation}%
where $N_{P}$ represents the number of volume cells comprising the primal lattice.
A similar Euler's polyhedron formula applies to the (closed, two-dimensional) boundary of the primal lattice
\begin{equation}
N_{V}^{b}-N_{E}^{b}=2-N_{F}^{b},  \label{Euler3b}
\end{equation}%

Combining Eq. (\ref%
{Euler31})\ \ and (\ref{Euler3b}), we have
\begin{equation}
\left( N_{E}^{{}}-N_{E}^{b}\right) -\left(
N_{V}^{{}}-N_{V}^{b}\right) =\left( N_{F}^{{}}-N_{F}^{b}\right)
-\left( N_{P}^{{}}-1\right). \label{Euler3}
\end{equation}%
From the Hodge decomposition (\ref{Hodge}), we see that $\Theta^d\left( E\right)$
is
\begin{eqnarray}
\Theta^d\left( E\right)  &=&N_{E}^{in}-N_{V}^{in}  \notag \\
&=&\left( N_{E}^{{}}-N_{E}^{b}\right) -\left(
N_{V}^{{}}-N_{V}^{b}\right). \label{Edof}
\end{eqnarray}%

Note that the divergence free condition $dB=0$
produces one constraint on the 2-form $B$ for each volume element. 
This constraint also span the whole lattice boundary. The total
number of the constrains for $B$ is therefore $\left(
N_{P}^{{}}-1\right) .$ Consequently,
we have
\begin{eqnarray}
\Theta^{d}\left( B\right)  &=&N_{F}^{in}-\left( N_{P}^{{}}-1\right)   \notag \\
&=&\left( N_{F}^{{}}-N_{F}^{b}\right) -\left( N_{P}^{{}}-1\right)
\label{Bdof}
\end{eqnarray}
so that
\begin{equation}
\Theta^{d}\left( B\right)=\Theta^{d}\left( E\right).
\end{equation}
This discussion can be generalized to lattices on non-contractible domains with any number of holes (genus), where
the identity $\Theta^{d}\left( B\right)=\Theta^{d}\left( E\right)$ is also satisfied~\cite{MineP25}. Moreover, from Hodge star isomorphism, we have $\Theta^{d}\left( D\right)=\Theta^{d}\left( E\right)$  and 
$\Theta^{d}\left( H\right)=\Theta^{d}\left( B\right)$.

In general, we can trace a direct correspondence between quantities in the Euler's polyhedron formula to the quantities in the Hodge decomposition formula. For example, each term in the two-dimensional Euler's formula $N_{E} =  N_{V} + \left( N_{F}-1\right) + g$
is associated to a corresponding term in 
$E = d\phi  +  \delta A  +  \chi$; that is, the number of edges $N_{E}$ corresponds to the dimension of the
space of lattice  $1$-forms $E$, which
is the sum of the
number of nodes $N_{V}$ (dimension of the space of discrete $0$-forms $\phi $%
), the number of faces $\left( N_{F}-1\right) $ (dimension of the
space of discrete $2$-forms $A$), and the number of holes $g$
(dimension of the space of harmonic forms $\chi $). A similar correspondence can be traced on a three-dimensional lattice~\cite{MineP25}. This
correspondence provides a physical picture to Euler's formula and a
geometric interpretation to the Hodge decomposition.

\vskip 18pt
\noindent {\bf Acknowledgments}
\vskip 12pt
The author thanks Weng C. Chew, Burkay Donderici, Bo He, Joonshik Kim, and David H. Adams for technical discussions. 

 \vskip 18pt
\noindent {\bf APPENDIX A: Differential forms and lattice fermions}
\vskip 12pt
Differential $p$-forms can be viewed as antisymmetric covariant tensor fields on rank $p$. Therefore, 
the ingredients discussed above are applicable to any antisymmetric tensor field theory, including (pure) non-Abelian theories~\cite{AdamsPRL97}. However, for (Dirac) fermion fields the situation is different and, at first, it would seem  unclear how differential forms could be used to describe spinors. Nevertheless, a useful connection can indeed be established~\cite{Montvay},\cite{BecherZPC82},\cite{Graf}. To briefly address this point, let us consider next the lattice transcription of the (one-flavor) Dirac equation.  Needless to say, the topic of lattice fermions is vast and we cannot do full justice to it here; we only focus here on the aspects more germane to our main discussion. In this Appendix, we work on Euclidean spacetime with $\hbar =c =1$ and adopt the repeated index summation convention with $\mu$, $\nu$ as coordinate indices, where $x$ is a point in four-dimensional space.

It is well known that fermion fields defy a lattice description with local coupling that gives the correct energy spectrum in the limit of zero lattice spacing and the correct chiral invariance ~\cite{AdamsPRD05}. This is formally stated by the no-go theorem of Nielsen-Ninomiya~\cite{Friedan} and is associated to the well-known `fermion-doubling' problem~\cite{Herbut}. A perhaps less known fact is that it is 
possible to arrive at a `geometrical' interpretation of the source of this difficulty by considering the 
`generalization' of the Dirac equation 
$(\gamma^\mu \partial_\mu +   m)\psi(x)=0$
given by the Dirac-K\"ahler equation
\begin{equation}
 (d-\delta)\Psi(x)=-m\Psi(x)
 \label{dk}
 \end{equation}
The square of the Dirac-K\"ahler operator can be viewed as the counterpart of the Dirac operator in the sense that 
\begin{equation}
(d-\delta)^2=-(d\delta + \delta d)=-\Box
\label{dksquared}
\end{equation}
recovers the Laplacian operator in the same fashion as the Dirac operator squared does, that is
$(\gamma^\mu \partial_\mu)^2=-\partial_\mu \partial^\mu=-\Box$, where $\gamma^\mu$ represents Euclidean gamma matrices.

The Dirac-K\"ahler equation admits a direct transcription on the lattice because both the exterior derivative $d$ and the
codifferential $\delta$ can be simply replaced by its lattice analogues, as discussed before.
However, for the Dirac equation the analogy has to further involve the relationship between the 4-component spinor field $\psi$ and the object $\Psi$. This relationship was first established in~\cite{BecherZPC82},\cite{Rabin} for hypercubic lattices and later extended to non-hypercubic lattices in~\cite{Gockeler},\cite{Raszillier}.
The analysis of~\cite{BecherZPC82} and~\cite{Rabin} has shown that $\Psi$ can be represented by a 16-component complex-valued inhomogeneous differential form:
\begin{equation}
\Psi(x)=\sum_{p=0}^{4} \alpha^{p}(x)
\end{equation}
where $\alpha^{0}(x)$ is a (1-component) scalar function of position or 0-form,
$
\alpha^1(x)=\alpha^1_\mu(x) dx^\mu$
is a (4-component) 1-form, and likewise for $p=2,3,4$ representing $2$-, $3$-, and $4$-forms with $6$-, $4$-, and $1$-components respectively. 
By employing the following Clifford algebra product
\begin{equation}
dx^{\mu} \vee dx^{\nu} = g^{\mu \nu} + dx^{\mu} \wedge dx^{\nu}
\end{equation}
as using the anti-commutative property of the exterior product $\wedge$, we have
\begin{equation}
dx^{\mu} \vee dx^{\nu} + dx^{\nu} \vee dx^{\mu} = 2 g^{\mu \nu}
\end{equation}
which exactly matches the anticommutator result of the $\gamma^\mu$ matrices, $\gamma^{\mu} \gamma^{\nu} + \gamma^{\nu} \gamma^{\mu} = 2 g^{\mu \nu}$. This suggests that $dx^{\mu}$ plays the role 
of the $\gamma^\mu$ matrix in the space of inhomogeneous differential forms with Clifford product~\cite{kanamori}, that is
\begin{equation}
\gamma^\mu \partial_\mu \mapsto dx^{\mu} \vee \partial_\mu 
\end{equation}
keeping in mind that while $\gamma^\mu \partial_\mu$ acts on spinors, whereas $dx^{\mu} \vee \partial_\mu =(d-\delta)$ acts on inhomogeneous differential forms.
This analysis leads to a `geometrical' interpretation of the popular Kogut-Susskind staggered lattice fermions~\cite{Susskind},\cite{MineP1} because the latter can be made identical to lattice Dirac-K\"ahler fermions after a simple relabeling of variables~\cite{Rabin}. 

The 16-component object $\Psi$ can be viewed as a $4\times 4$ matrix that produces a four-fold degeneracy with respect to the Dirac equation for $\psi$. This degeneracy is actually not a problem in the continuum because there is a well-defined procedure to extract the 4-components of $\psi$ from those of $\Psi$~\cite{BecherZPC82},\cite{Rabin} whereby the 16 scalar equations encoded by (\ref{dk}) all reduce to the same copy of the four equations encoded by
the standard Dirac equation. This procedure is performed by a set of `projection operators' that form a group~\cite{BecherZPC82},\cite{Benn}. On the lattice, however,
the operators $d$ and $\partial$, as well as $*$ (which plays a role on the space of inhomogeneous differential forms $\Psi$ analogous to that of $\gamma^5$ on the space of spinors $\psi$~\cite{Beauce}), behave in such a way that their action leads to lattice translations. This is because cochains with different $p$ necessarily live on different lattice elements and also because $*$ is a map between different lattice elements. As a consequence, the product operation of such `group' is not closed anymore. This nonclosure also stems from the fact that the lattice operators $d$ and $\delta$ do not satisfy Leibnitz's rule~\cite{kanamori}. Because of this, the degeneracy of the Dirac equation on the lattice is present at a more fundamental level and is harder to extricate using the Dirac-K\"ahler description than the analogous degeneracy in the continuum. 
In this regard, a new approach to identify the extraneous degrees of freedom away from the continuum was recently described in~\cite {AdamsPRL10}. In addition, a split-operator approach to solve Dirac equation based on the methods of characteristics that purports to avoid fermion doubling while maintaining chiral symmetry on the lattice was very recently put forth in~\cite{Fillion}. This approach preserves the 
linearity of the dispersion relation by a splitting of the original problem into a series of one-dimensional problems and the use of a upwind scheme with a Courant-Friedrichs-Lewy (CFL) number equal to one, which provides an exact time-evolution (i.e. with no numerical dispersion effects) along each reduced one-dimensional problem. The main (practical) obstacle in this case is the need to use very small lattice elements.

\vskip 18pt
\noindent {\bf APPENDIX B: Absorbing boundary conditions}
\vskip 12pt

In many wave scattering simulations, the presence of long-range interactions with slow (algebraic) decay, together with practical 
limitations in computer memory resources, 
implies that open-space problems necessitate the use of special techniques to suppress finite volume effects 
and emulate, for example, the Sommerfeld radiation condition at infinity.
Perfectly matched layers (PML) are absorbing boundary conditions
commonly used for this purpose~\cite{Berenger1994},\cite{Chew_Weedon1994},\cite{TeixeiraMGWL1997a},\cite{CollinoSIAM1998}. In the continuum limit, the PML provides a reflectionless
absorption of outgoing waves, in such a way that when the PML is used to truncate a computational lattice, finite volume effects such as spurious reflections from
the outer boundary are exponentially suppressed. 
When first introduced in the literature~\cite{Berenger1994}, 
the PML relied upon the use of matched artificial
electric and magnetic conductivities in Maxwell's equations and of a splitting of each vector field component into
two subcomponents. Because of this, the
resulting fields inside the PML layer are rendered `non-Maxwellian'.
The PML concept was later shown to be equivalent in the Fourier domain ($\partial_t \rightarrow -i\omega$) to a complex coordinate stretching of the coordinate space (or an analytic continuation to a complex-valued coordinate space)~\cite{Chew_Weedon1994},\cite{TeixeiraMGWL1997a},\cite{CollinoSIAM1998}
 and, as such, applicable to any linear wave phenomena.

Inside the PML, the (local) spatial coordinate $\zeta$ along the outward normal direction to each
lattice boundary point is complexified as
\begin{equation}
\zeta \rightarrow \tilde \zeta = \int_0^{\zeta} s_\zeta (\zeta') d\zeta'
\label{cmplxstr}
\end{equation}
where $s_\zeta$ is the so-called {\it complex stretching variable}
written as
$
s_\zeta(\zeta,\omega) = a_\zeta(\zeta) + i \Omega_\zeta(\zeta)/\omega
$
with $a_\zeta \ge 1$ and $\Omega_\zeta \ge 0$ (profile functions). 
The first inequality ensures that evanescent waves will have a faster exponential decay
in the PML region, and the second inequality ensures that propagating waves will
decay exponentially along $\zeta$ inside the PML. 
As opposed to some other lattice truncation techniques, 
the PML preserves the locality of the underlying differential operators 
and hence retains the sparsity of the formulation.

For Maxwell's equations, the PML can also be effected  
by means of artificial material tensors (Maxwellian PML)
~\cite{Sacks1995}. In three-dimensions, the 
Maxwellian PML can be represented as a media with anisotropic
permittivity and permeability tensors exhibiting stratification along the normal to the 
boundary $S$ that parametrizes the lattice truncation boundary. The PML tensors properties depend on the local geometry via the two principal curvatures of $S$~\cite{TeixeiraMGWL1997b},\cite{TeixeiraMOTL1998},\cite{dondericiAP08}. The boundary surface $S$ is assumed (constructed) as doubly differentiable with non-negative radii of curvature, otherwise dynamic instabilities ensue during a marching-on-time evolution ~\cite{MineP18}. 

From (\ref{cmplxstr}), the PML also admits a straightforward interpretation 
as a complexification of the metric~\cite{MineP17},\cite{MineP19}. As a result, 
the use of differential forms readily unifies the Maxwellian and non-Maxwellian PML formulations because the metric is explicitly factored out into the Hodge star operators---any transformation the metric corresponds, dually, to a transformation on the Hodge star operators that can be 
mimicked by modified constitutive relations~\cite{MineP15}. 
In the differential forms framework, the PML is obtained by a mapping on the Hodge star operators:
$
\star_\epsilon \rightarrow \tilde \star_\epsilon 
$ and
$ 
\star_{\mu^{-1}} \rightarrow \tilde \star_{\mu^{-1}}$
induced by the complexification of the metric. The resulting differential forms inside the PML, 
$\tilde E, \tilde D, \tilde H, \tilde B$ therefore obey `modified' Hodge
relations
$\tilde D = \tilde \star_\epsilon  \tilde E$ and $
\tilde B =  \tilde \star_{\mu^{-1}}  \tilde H,$
but identical pre-metric equations (\ref{faraday}) and (\ref{ampere}). In other words, (\ref{faraday}) and (\ref{ampere}) are invariant under the transformation~(\ref{cmplxstr})~\cite{MineP17},\cite{MineP19}.

\vskip 18pt

\noindent {\bf APPENDIX C: Implementation of space charge effects}

\vskip 12pt

In many applications related to plasma physics or electronic devices, it is necessary to include space charges (uncompensated charge effects) into lattice models of macroscopic Maxwell's equations. This is typically done by representing the charged plasma media using particle-in-cell (PIC) methods that track the individual particles on the lattice~\cite{Hockney},\cite{esirkepov},\cite{ok06b}. The field/charge interaction is then modeled by ($i$) interpolating lattice fields (cochains) to particle positions (gather step), ($ii$) advancing particle positions and velocities in time using equations of motion, and ($iii$) interpolating back charge densities and currents onto the lattice as cochains (scatter step). In general, the `particles' do not need to be actual individual particles, but can be a collection thereof (`macro-particles').
To put it simply, incorporation of space charges requires two extra steps during the field update in any marching-on-time algorithm, which transfer information from the instantaneous field distribution to the particle kinematic update and vice-versa. Conventionally, this information transfer relies on spatial interpolations that often violate the charge continuity equation and, as a result, lead to spurious charge deposition on the lattice nodes. On regular lattices, this problem can be corrected, for example, using approaches that either subtract a static solution (charges) from the electric field solution (Boris/DADI correction) or directly subtract the residual error on the Gauss' law (Langdon-Marder correction) at each time step~\cite{Mardahl}. On irregular lattices, additional degrees of freedom can be added as coupled elliptical constraints to produce a augmented Lagrange multiplier system~\cite{Assous}. All these approaches necessitate changes on the original equations, while still allowing for small violations on charge conservation.
In contrast, Whitney forms provide a direct route to construct gather and scatter steps that satisfy charge conservation exactly even on unstructured lattices~\cite{candel09},\cite{squire12}, as explained next. To conform to the vast majority of the plasma and electronic devices literature, we once more restrict ourselves here to the 3+1 setting (although a four-dimensional analysis in Minkowski space would have provided a more succinct discussion).

For the gather step, Whitney forms can be used to directly compute (interpolate) the fields at any location from the knowledge of its cochain values, such as in
(\ref{exp1}) and (\ref{exp2}) for example.
For the scatter step, charge movement can be modeled
as the Hodge-dual of the current 2-form $J$, that is, as the 1-form $\star J$ which can be expanded in terms of Whitney 1-forms on the primal lattice. Here, $\star$ represents again the spatial Hodge star in three-dimensions distilled from macroscopic constitutive properties. The Hodge-dual current associated to an individual point charge can be expressed as $\star J = q v^{\flat}$, where 
$q$ is the charge value, $v$ is the associated velocity vector, and $\flat$ is the `flat' operator or index-lowering canonical isomorphism that maps a vector to a 1-form, given by the Euclidean metric. 
Similarly, point charges can be encoded as the Hodge-dual of the charge density 3-form $\rho$, that is, as the 0-form $\star \rho$, which can be expanded in terms of Whitney 0-forms on the primal lattice.
These two Whitney maps are linked in such a way that the rate of change on the value of the 0-cochain representing $\star \rho$ at a node is associated to the presence of a 1-cochain representing $\star J$ along the edges that touch that particular node, leading to exact charge conservation at the discrete level. To show this, consider for simplicity the two-dimensional case of a point charge $q$ moving from point $x^{(s)}$ to point $x^{(f)}$ during a time interval $\tau$ inside a triangular cell with nodes $\sigma_{0,0}$, $\sigma_{0,1}$, and $\sigma_{0,2}$, or simply $0$, $1$, and $2$.
At any point $x$ inside this cell, the 0-form $\star \rho$ can be scattered to these three adjacent nodes via 
\begin{equation}
\star \rho = q \sum_{i=1}^3 \left<x, \omega^0_i \right> \omega^0_i
  \label{scatter1}
  \end{equation}
where we are again using the short-hand $\omega^0[\sigma_{0,i}]=\omega^0_i$, and the brackets represent the pairing expressed by (\ref{pairing}). In this case, $p=0$ and the pairing integral in (\ref{pairing}) reduces to a function evaluation at a point. 
Since Whitney 0-forms are equal to the barycentric coordinates associated of a given node, that is  $\left<x, \omega^0_i \right>=\lambda_i(x)$, we have the scattered charge $q \lambda^s_i \doteq q \lambda_i(x^{(s)})$ on node $i$ for a charge $q$ at $x^{(s)}$, and, similarly,
the scattered charge 
$q \, \lambda^f_i$ on node $i$ for a charge $q$ at $x^{(f)}$.
The rate of scattered charge variation on a given node $i$
is therefore equal to $\dot{q}(\lambda^f_i -\lambda^s_i)$, where $\dot{q}=q/\tau$.

During $\tau$, the particle travels through a path $\ell$ from
$x^{(s)}$ to $x^{(f)}$, and the corresponding $\star J$ can be expanded as a sum of  
Whitney 1-forms $\omega^1_{\overline{ij}}$ associated to the three adjacent edges $\overline{ij}=\overline{01},\overline{12},\overline{20}$, that is
\begin{equation}
\star J = \dot{q}\sum_{\overline{ij}} \left<\ell,\omega^1_{\overline{ij}}\right>\omega^1_{\overline{ij}}  
\label{scatter2}
\end{equation}
The coefficients $\left<\ell,\omega^1_{\overline{ij}}\right>$ represent the (oriented) current flow along the associated oriented edge, that is, the cochain representation of $\star J$ along edge $\overline{ij}$. Using (\ref{Whitney}),
the sum of the total current magnitude scattered along edges $\overline{01}$ and $\overline{20}$ that flows {\it into} node $0$ is therefore
\begin{equation}
 \dot{q}\left( -\left<\ell,\omega^1_{\overline{01}}\right>+\left<\ell,\omega^1_{\overline{20}}\right> \right)=
\dot{q} \int_\ell \left( - \omega^1_{\overline{01}} + \omega^1_{\overline{20}}\right)
\end{equation}
Using $\omega^1_{\overline{ij}}=\lambda_i d \lambda_j - \lambda_j d \lambda_i$
and $\lambda_1 + \lambda_2 + \lambda_3 = 1$,
the above reduces to
\begin{equation}
 \dot{q} \int_\ell d\lambda_0=
\dot{q} (\lambda^f_0 -\lambda^s_0)
\end{equation}
which exactly matches the rate of scattered charge variation on node $0$ obtained before. It is clear that similar equalities hold for nodes 1 and 2. More fundamentally, these equalities are a direct consequence of the structural property (\ref{struc2}).

\vskip 18pt

\noindent {\bf APPENDIX D: Classification of inconsistencies in na\"ive discretizations}

\vskip 12pt

We provide below a rough classification scheme of inconsistencies arising from na\"ive
discretizations of the differential calculus on irregular lattices.

\vskip 6pt

\noindent {\it (a) Pre-metric inconsistencies of first kind:}
We call pre-metric inconsistencies of the first kind those that are related to the primal or dual lattices taken as separate objects and that occur when the discretization violates one or more properties of the continuum theory that is invariant under homeomorphisms---for example, conservations laws that relate a quantity on a region $S$ with an associated
quantity on the boundary of the region, $\partial S$ (a topological invariant). 
Perhaps the most illustrative example is violation of `divergence-free' conditions caused by
improper construction of incidence matrices, 
whereby the nilpotency of the (adjoint) boundary operator, $\partial \circ \partial = 0$, is not observed. 
This implies, in a dual fashion, that the identity
$d^2 = 0$ is violated~\cite{MineP16}. Stated in another way,
the exact sequence property of the underlying de Rham differential complex is violated~\cite{Arnold2002}.
In practical terms, this leads to the appearance spurious charges and/or spurious (`ghost') modes. 
As the classification suggests, 
these properties are not related to metric aspects of the lattice, but only to its ``topological aspects'' that is, on how discrete calculus operators are defined vis-\`{a}-vis the lattice element {\it connectivity}. 
In more mathematical terms, one can say that the structure of the (co)homology groups of the continuum manifold is not correctly captured by the cell complex (lattice).
We stress again that, given any dual lattice construction, pre-metric inconsistencies of the first kind are associated to the primal or dual lattice taken separately, and not necessarily on how they intertwine.

\vskip 6pt

\noindent {\it (b) Pre-metric inconsistencies of second kind:}
The second type of pre-metric inconsistency is associated to the breaking of some {\it discrete} symmetry of the Lagrangian. In mathematical terms, this type of inconsistency can occur when the bijective correspondence between
$p$-cells of the primal lattice and $(n-p)$-cells of the dual lattice (an expression of Poincar\'e duality at the level of cellular homology~\cite{munkres}, up to boundary terms) is violated. This is typified by `nonreciprocal' constructions 
of derivative operators, where the boundary operator effecting the spatial derivation 
on the primal lattice $K$ is not the dual adjoint 
(or the incidence matrix transpose)  
of the boundary operator on the dual lattice $\mathcal K$: for example, the identity
${\tilde C}_{ij}^p = C_{ji}^{n-1-p}$ (up to boundary terms) used to obtain eq.~(\ref{primallat}) is violated.
One basic consequence of this violation is that the resulting discrete equations break time-reversal symmetry. Consequently, the numerical solutions 
will violate energy conservation and produce either artificial dissipation or late-time instabilities~\cite{MineP16}.
Many algorithms developed over the years 
for hyperbolic partial differential equations do indeed violate these properties: they are dissipative and cannot be used for long integration times~\cite{ChevalierAP97},\cite{WhiteMTT01}. 
It should be noted at this point that lattice field theories invariably break Lorentz covariance and many of the {\it continuum} Lagrangian symmetries and, as a result, violate conservation laws (currents) by virtue of Noether's theorem. For example, angular momentum conservation does not hold exactly on the lattice because
of the lack of continuous rotational symmetry (note that discrete rotational symmetries can still be present). However, this latter type of symmetry breaking is of a fundamentally
different nature because it is `controllable', i.e. their effect on the computed solutions is made arbitrarily small in the continuum limit. More importantly,
discrete transcriptions of the Noether's theorem can be constructed for Lagrangian symmetries on a lattice~\cite{SorkinJMP75},\cite{christ12}, to yield {\it exact}
conservation laws of (properly defined) quantities such as discrete energy and discrete momentum~\cite{Chew}. 

\vskip 6pt

\noindent {\it (c) Hodge-star inconsistencies:}
In the third type of inconsistency, we include those that
arise in connection with metric properties of the lattice. Because the
metric is entirely encoded in the Hodge-star operators~\cite{MineP16},\cite{BossavitPIER32},\cite{HiptmairNM01}, such inconsistencies can be simply understood as
inconsistencies on the construction of discrete Hodge-star operators (or their procedural analogues). 
For example, it is not uncommon for na\"ive discretizations in irregular lattices to yield asymmetric discrete Hodge operators, as noted in~\cite{WeilandIJNM96},\cite{RailtonEL97}. 
Even if symmetry is observed, non positive definiteness 
might ensue that is often associated with portions of the lattice with highly skewed or obtuse cells~\cite{Schuhmann98}. 
Lack of either of these properties lead to unconditional instabilities that destroy marching-on-time solutions~\cite{MineP16}. When very long integration times are needed, 
asymmetry in the discrete Hodge matrices
can be a problem even if produced at the level of machine rounding-off errors.

\vskip 18pt

\noindent {\bf APPENDIX E: Overview of related discretization approaches}
\vskip 12pt
We outline below some discretization programs that rely, one way or another, on tenets exposed above. This delineation is mostly informed mostly by applications related to electrodynamics and not too sharp as the programs share much in common.
\vskip 6pt
\noindent {\it (a) Finite-difference time-domain method}: In cubical lattices, the (lowest-order) Whitney forms can be represented by means of a product of pulse and `rooftop' functions on the three Cartesian coordinates~\cite{chilton2008}. This choice, together with the use of low-order quadrature rules to compute the Hodge star integrals in (\ref{hodge1m}) and (\ref{hodge2m}), leads to {\it diagonal} matrices $[\star_\epsilon]$, $[\star_{\mu^{-1}}]$, and, consequently, also diagonal $[\star_\epsilon]^{-1}$, $[\star_{\mu^{-1}}]^{-1}$ and sparse 
$[\Upsilon]$ so that an ultra-local equation results for (\ref{far2}). In this fashion, one obtains a `matrix-free' algorithm where no linear algebra is needed during a marching-on-time solution for the fields. 
This prescription recovers Yee's finite-difference time-domain scheme~\cite{YeeAP69},\cite{Taflove95},\cite{YeeAP97}.
Conventional FDTD adopts the simplest explicit, energy-conserving (symplectic) time-discretization for eqs.~(\ref{far1}) and (\ref{far2}), which can be constructed by staggering 
  the electric and magnetic fields in time and replacing time derivatives by central differences. Staggering in both space and time is consistent with the presence of two staggered hypercubical spacetime lattices~\cite{TontiPIER01},\cite{MattiussiPIER01}. The staggering in time also provides a $O(\Delta t^2)$ truncation error. 

 \vskip 6pt
 
\noindent {\it (b) Finite integration technique}: The finite integration technique (FIT)~\cite{Weiland84},\cite{Schuhmann00},\cite{codecasa04} is closely related to FDTD, the main distinction being that, assuming piecewise constant fields over each cell, the latter is equivalent to applying the (discrete version) of the generalized Stokes' theorem to the cochains in
(\ref{far1}) and (\ref{ampx}).
Another difference is that the incidence matrices and material (Hodge star) matrices are treated separately in FIT, in a manner akin to that exposed in Sections III and IV.
Like FDTD, FIT is based on dual staggered lattices and, for cubical lattices, it turns out that the lowest-order numerical implementation of FIT
is equivalent to the lowest-order FDTD. The spatial operators in FIT can all be viewed as discrete incarnations of the exterior derivative for the various $p$, and as such, the exact sequence property of the underlying de Rham complex is automatically enforced by construction~\cite{BossavitIEE88}. Historically, FIT generalizations to irregular lattices have relied on the use of either projection operators~\cite{Schuhmann98} or Whitney forms~\cite{schuhmann02} to construct discrete versions of the Hodge star operators (or their procedural equivalents); however, these generalizations do not necessarily recover the specific form of the discrete Hodge matrix elements expressed in 
(\ref{hodge1m}) and (\ref{hodge2m}). 

\vskip 6pt

\noindent {\it (c) Cell method}: Another related discretization program, based on general principles originally put forth in~\cite{TontiPIER01},\cite{TontiRAL72},\cite{tonti76}, is the Cell method~\cite{bullo04},\cite{alotto06},\cite{bullo06},\cite{alotto08},\cite{alotto10},\cite{codecasa}. Even though this program does not rely on Whitney forms for constructing discrete Hodge star operators (other geometrically-based constructions are used instead), it is nevertheless still based upon the use of `domain-integrated' discrete variables that conform to the notion of discrete differential forms or cochains of various degrees and, as such, it is naturally suited for irregular lattices. The Cell method also employs
metric-free discrete operators that satisfy the exactness property of the de Rham complex and make explicit use of a dual lattice (but not necessarily barycentric) motivated by the notion of inner and outer orientations. The relationships between the various discrete operators and `domain-integrated' field quantities (cochains) in the Cell method are built into general classification diagrams referred to as `Tonti diagrams' that reproduce correct 
commuting diagram properties of the underlying operators~\cite{tonti76},\cite{TontiPIER01}.

\vskip 6pt

\noindent {\it (d) Mimetic finite-differences}: `Mimetic' finite-difference methods, originally developed for non-orthogonal hexahedral structured lattices (`tensor-product grids') and later extended for irregular and polyhedral lattices~\cite{SteinbergJCP95},\cite{ShashkovJCP99},\cite{ShashkovSIAM99},\cite{ShashkovANM97},\cite{ShashkovPIER01},\cite{CastilloANM02},\cite{lipnikov06},\cite{brezzi10},\cite{robidoux11},\cite{lipnikov11} also share many of the properties exposed above. The thrust here is towards the construction of discrete versions of the differential operators divergence, gradient, and curl of vector calculus having `compatible' (in the sense of the exactness property of the underlying de Rham complex) 
domains and ranges and such that the resulting discrete equations exactly satisfy discrete conservation laws. In three dimensions, this naturally leads to the definition of three `natural' operators and three `adjoint' operators that can be associated with exterior derivative $d$ and the codifferential $\delta$, respectively, for $p=1,2,3$ (although the exterior calculus terminology is often not used explicitly in this context). 
In mimetic finite-differences, the discrete analogues of the codifferential operator $\delta$
are full matrices, and the matrix-free character of FDTD is lacking even on orthogonal lattices. 
A very thorough, historical review of mimetic finite-difference is provided in~\cite{Lipnikov}.

\vskip 6pt

\noindent {\it (e) Compatible discretizations and finite element exterior calculus}:
In recent years, much attention has been devoted to the development of `compatible discretizations,' an umbrella term used to denote spatial discretizations of partial differential equations seeking to provide finite element spaces that reproduce the exactness of the underlying de Rham complex (or the correct cohomology in topologically nontrivial domains)~\cite{arnold02},\cite{arnold06a},\cite{white06},\cite{bochev06},\cite{boffi07},\cite{Bochev12}. In this program, Whitney forms play a role of providing `conforming' vector-valued functional (finite element) spaces of Sobolev-type. Specifically, Whitney 1-forms recover the space of `Nedelec edge-elements' or curl-conforming Sobolev space ${\bf H}(\text{curl},\Omega)$ ~\cite{nedelec80} and Whitney 2-forms recover the space of `Raviart-Thomas elements'
or div-conforming Sobolev space ${\bf H}(\text{div},\Omega)$~\cite{hiptmairMC99}.
In this regard, a relatively new advance here has been the development of new finite element spaces, beyond those provided by Whitney forms, based on the Koszul complex~\cite{Guil}. The latter is key for the stable discretization of elastodynamics~\cite{arnold06b}. Another recent approach aimed at the stable discretization of elastodynamics 
is described in~\cite{Yavari08}.
The link between stability conditions of some mixed finite element methods~\cite{nedelec80} and the complex of Whitney forms has a long history in the context of electrodynamics~\cite{BossavitIEE88},\cite{Bossavitchap},\cite{Bossavit98},\cite{BossavitEJM91},\cite{KettunenMAGN98},\cite{KettunenMAGN99},\cite{MineP27},\cite{KotiugaJAP93},\cite{kangas07},\cite{wong95},\cite{feliziani98},\cite{castillo04},\cite{rieben05}.

\vskip 6pt

\noindent {\it (f) Discrete exterior calculus}: The `discrete exterior calculus' (DEC) is yet another discretization program aimed at developing {\it ab initio} consistent discrete models to describe field theories~\cite{squire12},\cite{Desbrun03},\cite{Hirani03},\cite{Desbrun05},\cite{Gillette09},\cite{perot}. This program recognizes the role played by discrete differential forms to capture and the need for dual lattices to capture the correct physics. Note that DEC has focused on the use of a circumcentric dual as opposed to a barycentric dual~\cite{Hirani03},\cite{Desbrun05} (despite the fact that the former does not admit a metric-free construction) and does not emphasize the role of Whitney forms. DEC also recognizes the need to address group-valued differential forms, as well as the mathematical objects that exist on the dual-bundle space together with the associated operators (such as contractions and Lie derivatives), in connection  
to discrete problems in mechanics, optimal control, and computer vision/graphics~\cite{Desbrun03}. A recent discussion on obstacles associated with some of the DEC underpinnings is provided in~\cite{Kotiuga08}


\bibliographystyle{ieeetr}
\bibliography{newbib,doeprop,fdtd,fem,forms,subgrid,mypapers,mimetic,pml}

\begin{thebibliography}{100}

\bibitem{Montvay}
I.~Montvay and G.~Munster, {\em Quantum Fields on a Lattice}.
\newblock Cambridge, U.K.: Cambridge University Press, 1997.

\bibitem{Zee}
A.~Zee, {\em Quantum Field Theory in a Nutshell}.
\newblock Princeton, NJ: Princeton University Press, 2003.

\bibitem{Chew}
W.~C. Chew, ``{Electromagnetic field theory on a lattice},'' {\em J. Appl.
  Phys.}, vol.~75, pp.~4843--4850, 1994.

\bibitem{sanmartin}
L.~S. Martin and Y.~Oono, ``Physics-motivated numerical solvers for partial
  differential equations,'' {\em Phys. Rev. E}, vol.~57, pp.~4795--4810, 1998.

\bibitem{Bretones}
M.~A.~H. Lopez, S.~G. García, A.~R. Bretones, and R.~G. Martin, ``Simulation of
  the transient response of objects buried in dispersive media,'' {\em
  Ultrawideband Short-Pulse Electromagnetics}, vol.~5, Kluwer Academic Press,
  2000.

\bibitem{TeixeiraAP08}
F.~L. Teixeira, ``{Time-domain finite-difference and finite-element methods for
  Maxwell equations in complex media},'' {\em IEEE Trans. Antennas Propagat.},
  vol.~56, pp.~2150--2166, 2008.

\bibitem{Christ}
N.H.Christ, R.~Friedberg, and T.~Lee, ``Gauge theory on a random lattice,''
  {\em Nucl. Phys. B}, vol.~210, pp.~310--336, 1982.

\bibitem{Bolander}
J.~E. Bolander and N.~Sukumar, ``Irregular lattice model for quasistatic crack
  propagation,'' {\em Phys. Rev. B}, vol.~71, p.~094106, 2005.

\bibitem{Drouffe}
J.~M. Drouffe and K.~J.~M. Moriarty, ``{U(2) four-dimensional simplicial
  lattice gauge theory},'' {\em Z. Phys. C}, vol.~24, pp.~395--403, 1984.

\bibitem{Gockeler}
M.~G\"ockeler, ``{Dirac-K\"ahler fields and lattice shape dependence of fermion
  flavour},'' {\em Z. Phys. C}, vol.~18, pp.~323--326, 1983.

\bibitem{KomorowskiBAPS75}
J.~Komorowski, ``{On finite-dimensional approximations of the exterior
  differential, codifferential, and Laplacian on a Riemannian manifold},'' {\em
  Bull. L'Acad. Pol. Sci}, vol.~23, no.~9, pp.~999--1005, 1975.

\bibitem{DodziukAJM76}
J.~Dodziuk, ``{Finite-difference approach to the Hodge theory of harmonic
  forms},'' {\em Am. J. Math.}, vol.~98, no.~1, pp.~79--104, 1976.

\bibitem{SorkinJMP75}
R.~Sorkin, ``{The electromagnetic field on a simplicial net},'' {\em J. Math.
  Phys.}, vol.~16, no.~12, pp.~2432--2440, 1975.

\bibitem{weingarten77}
D.~Weingarten, ``Geometric formulation of electrodynamics and general
  relativity in discrete space-time,'' {\em J. Math. Phys.}, vol.~18,
  pp.~165--170, 1977.

\bibitem{MullerAM78}
W.~Muller, ``{Analytic torsion and R-torsion of Riemannian manifolds},'' {\em
  Adv. Math.}, vol.~28, pp.~233--305, 1978.

\bibitem{BecherZPC82}
P.~Becher and H.~Joos, ``{The Dirac-Kahler equation and fermions on the
  lattice},'' {\em Z. Phys. C}, vol.~15, pp.~343--365, 1982.

\bibitem{Rabin}
J.~M. Rabin, ``Homology theory of lattice fermion doubling,'' {\em Nucl.
  Phys.}, vol.~B201, pp.~315--332, 1982.

\bibitem{Bossavit98}
A.~Bossavit, {\em Computational Electromagnetism}.
\newblock New York: Academic Press, 1998.

\bibitem{BossavitEJM91}
A.~Bossavit, ``{Differential forms and the computation of fields and forces in
  electromagnetism},'' {\em Eur. J. Mech. B, Fluids}, vol.~10, no.~5,
  pp.~474--488, 1991.

\bibitem{AdamsArXiv}
D.~H. Adams, ``{R-torsion and linking numbers from simplicial Abelian gauge
  theories}.'' arXiv:hep-th/9612009.

\bibitem{MattiussiJCP97}
C.~Mattiussi, ``{An analysis of finite volume, finite element, and finite
  difference methods using some concepts from algebraic topology},'' {\em J.
  Comp. Phys.}, vol.~133, pp.~289--309, 1997.

\bibitem{KettunenMAGN98}
L.~Kettunen, K.~Forsman, and A.~Bossavit, ``{Discrete spaces for div and
  curl-free fields},'' {\em IEEE Trans. Magn.}, vol.~34, pp.~2551--2554, 1998.

\bibitem{MineP16}
F.~L. Teixeira and W.~C. Chew, ``{Lattice electromagnetic theory from a
  topological viewpoint},'' {\em J. Math. Phys.}, vol.~40, no.~1, pp.~169--187,
  1999.

\bibitem{KettunenMAGN99}
T.~Tarhasaari, L.~Kettunen, and A.~Bossavit, ``{Some realizations of a discrete
  Hodge operator: A reinterpretation of finite element techniques},'' {\em IEEE
  Trans. Magn.}, vol.~35, pp.~1494--1497, 1999.

\bibitem{SenAdamsPRE00}
S.~Sen, S.~Sen, J.~C. Sexton, and D.~H. Adams, ``{Geometric discretization
  scheme applied to the Abelian Chern-Simons theory},'' {\em Phys. Rev. E},
  vol.~61, no.~3, pp.~3174--3185, 2000.

\bibitem{ShapiroMCS00}
J.~A. Chard and V.~Shapiro, ``{A multivector data structure for differential
  forms and equations},'' {\em Math. Comp. Simulat.}, vol.~54, pp.~33--64,
  2000.

\bibitem{KotiugaPIER01}
P.~W. Gross and P.~R. Kotiuga, ``{Data structures for geometric and topological
  aspects of finite element algorithms},'' in {\em Geometric Methods in
  Computational Electromagnetics, PIER 32} (F.~L. Teixeira, ed.), pp.~151--169,
  Cambridge, Mass.: EMW Publishing, 2001.

\bibitem{TeixeiraPIER01}
F.~L. Teixeira, ``{Geometrical aspects of the simplicial discretization of
  Maxwell's equations},'' in {\em Geometric Methods in Computational
  Electromagnetics, PIER 32} (F.~L. Teixeira, ed.), pp.~171--188, Cambridge,
  Mass.: EMW Publishing, 2001.

\bibitem{KettunenPIER01}
T.~Tarhasaari and L.~Kettunen, ``{Topological approach to computational
  electromagnetism},'' in {\em Geometric Methods in Computational
  Electromagnetics, PIER 32} (F.~L. Teixeira, ed.), pp.~189--206, Cambridge,
  Mass.: EMW Publishing, 2001.

\bibitem{MineP31}
J.~Kim and F.~L. Teixeira, ``{Parallel and explicit finite-element time-domain
  method for Maxwell's equations},'' {\em IEEE Trans. Antennas Propagat.},
  vol.~59, no.~6, pp.~2350--2356, 2011.

\bibitem{Wise}
D.~K. Wise, ``{p-form electromagnetism on discrete spacetimes},'' {\em Class.
  Quantum Grav.}, vol.~23, pp.~5129--5176, 2006.

\bibitem{Schwarz94}
A.~S. Schwarz, {\em Topology for Physicists}.
\newblock New York: Springer-Verlag, 1994.

\bibitem{MineP25}
B.~He and F.~L. Teixeira, ``{On the degrees of freedom of lattice
  electrodynamics},'' {\em Phys. Lett. A}, vol.~336, no.~1, pp.~1--7, 2005.

\bibitem{MineP27}
B.~He and F.~L. Teixeira, ``{Mixed E-B finite elements for solving 1-D, 2-D,
  and 3-D time-harmonic Maxwell curl equations},'' {\em IEEE Microw. Wireless
  Comp. Lett.}, vol.~17, no.~5, pp.~313--315, 2007.

\bibitem{Whitney57}
H.~Whitney, {\em Geometric Integration Theory}.
\newblock Princeton, NJ: Princeton University Pres, 1957.

\bibitem{Misner}
C.~W. Misner, K.~S. Thorne, and J.~A. Wheeler, {\em Gravitation}.
\newblock New York: Freeman and Co., 1973.

\bibitem{DeschampsIEEE82}
G.~A. Deschamps, ``{Electromagnetics and differential forms},'' {\em Proc.
  IEEE}, vol.~69, pp.~676--696, 1982.

\bibitem{KotiugaJAP93}
P.~R. Kotiuga, ``{Metric dependent aspects of inverse problems and functionals
  based on helicity},'' {\em J. Appl. Phys.}, vol.~73, pp.~5437--5439, 1993.

\bibitem{MineP15}
F.~L. Teixeira and W.~C. Chew, ``{Unified analysis of perfectly matched layers
  using differential forms},'' {\em Microw. Opt. Technol. Lett.}, vol.~20,
  no.~2, pp.~124--126, 1999.

\bibitem{MineP17}
F.~L. Teixeira and W.~C. Chew, ``{Differential forms, metrics, and the
  reflectionless absorption of electromagnetic waves},'' {\em J. Electromagn.
  Waves Applicat.}, vol.~13, no.~5, pp.~665--686, 1999.

\bibitem{MineP29}
F.~L. Teixeira, ``{Differential form approach to the analysis of
  electromagnetic cloaking and masking},'' {\em Microw. Opt. Technol. Lett.},
  vol.~49, no.~8, pp.~2051--2053, 2007.

\bibitem{Guth}
A.~H. Guth, ``Existence proof od a nonconfining phase in four-dimensional
  {U}(1) lattice field theory,'' {\em Physical Review D}, vol.~21, no.~8,
  pp.~2291--2307, 1980.

\bibitem{Kheyfets}
A.~Kheyfets and W.~A. Miller, ``The boundary of a boundary in field theories
  and the issue of austerity of the laws of physics,'' {\em J. Math. Phys.},
  vol.~32, no.~11, pp.~3168--3175, 1991.

\bibitem{HiptmairNM01}
R.~Hiptmair, ``{Discrete Hodge operators},'' {\em Numer. Math.}, vol.~90,
  pp.~265--289, 2001.

\bibitem{MineP26}
B.~He and F.~L. Teixeira, ``{Geometric finite element discretization of Maxwell
  equations in primal and dual spaces},'' {\em Phys. Lett. A}, vol.~349,
  no.~1-4, pp.~1--14, 2006.

\bibitem{MineP28}
B.~He and F.~L. Teixeira, ``{Differential forms, Galerkin duality, and sparse
  inverse approximations in finite element solutions of Maxwell equations},''
  {\em IEEE Trans. Antennas Propagat.}, vol.~55, no.~5, pp.~1359--1368, 2007.

\bibitem{MineP30}
B.~Donderici and F.~L. Teixeira, ``{Mixed finite-element time-domain method for
  transient Maxwell equations in doubly dispersive media},'' {\em IEEE Trans.
  Microwave Theory Tech.}, vol.~56, no.~1, pp.~113--120, 2008.

\bibitem{burke}
W.~L. Burke, {\em Applied Differential Geometry}.
\newblock Cambridge University Press, 1985.

\bibitem{tonti76}
E.~Tonti, ``The reason for analogies between physical theories,'' {\em Appl.
  Math. Model.}, vol.~1, pp.~37--50, 1976.

\bibitem{TontiPIER01}
E.~Tonti, ``{Finite formulation of the electromagnetic field},'' in {\em
  Geometric Methods in Computational Electromagnetics, PIER 32} (F.~L.
  Teixeira, ed.), pp.~1--44, Cambridge, Mass.: EMW Publishing, 2001.

\bibitem{TontiRAL72}
E.~Tonti, ``{On the mathematical structure of a large class of physical
  theories},'' {\em Rend. Accad. Lincei}, vol.~52, pp.~48--56, 1972.

\bibitem{YeeAP69}
K.~S. Yee, ``{Numerical solution of initial boundary value problems involving
  Maxwell'sequation is isotropic media},'' {\em IEEE Trans. Antennas
  Propagat.}, vol.~14, no.~3, pp.~302--307, 1969.

\bibitem{Taflove95}
A.~Taflove, {\em Computational Electrodynamics: The Finite-Difference
  Time-Domain Method}.
\newblock Norwood, MA: Artech House, 1995.

\bibitem{NicolaidesSIAM97}
R.~A. Nicolaides and X.~Wu, ``{Covolume solutions of three-dimensional div-curl
  equations},'' {\em SIAM J. Numer. Anal.}, vol.~34, no.~6, pp.~2195--2203,
  1997.

\bibitem{codecasa07}
L.~Codecasa, R.~Specogna, and F.~Trevisan, ``Symmetric positive-definite
  constitutive matrices for discrete eddy-current problems,'' {\em IEEE Trans.
  Magn.}, vol.~43, pp.~510--515, 2007.

\bibitem{Auchmann}
B.~Auchmann and S.~Kurz, ``{A geometrically defined discrete Hodge operator on
  simplicial cells},'' {\em IEEE Trans. Magn.}, vol.~42, pp.~643--646, 2006.

\bibitem{BossavitIEE88}
A.~Bossavit, ``{Whitney forms: A new class of finite elements for
  three-dimensional computations in electromagnetics},'' {\em IEE Proc. A},
  vol.~135, pp.~493--500, 1988.

\bibitem{kotiugabook}
P.~W. Gross and P.~R. Kotiuga, {\em Electromagnetic Theory and Computation: A
  Topological Approach}.
\newblock Cambridge University Press, 2004.

\bibitem{bossavit05}
A.~Bossavit, ``{Discretization of electromagnetic problems: The 'generalized
  finite-differences approach'},'' in {\em Numerical Methods in
  Electromagnetism (Handbook of Numerical Analysis, Vol. 13)}, pp.~105--197,
  Amsterdam: Elsevier, 2005.

\bibitem{bohethesis}
B.~He, {\em Compatible Discretizations of Maxwell Equations}.
\newblock PhD thesis, The Ohio State University, 2006.

\bibitem{HiptmairPIER01}
R.~Hiptmair, ``{Higher order Whitney forms},'' in {\em Geometric Methods in
  Computational Electromagnetics, PIER 32} (F.~L. Teixeira, ed.), pp.~271--299,
  Cambridge, Mass.: EMW Publishing, 2001.

\bibitem{rapetti}
F.~Rapetti and A.~Bossavit, ``Whitney forms of higher degree,'' {\em SIAM J.
  Numer. Anal.}, vol.~47, pp.~2369--2386, 2009.

\bibitem{kangas07}
J.~Kangas, T.~Tarhasaari, and L.~Kettunen, ``{Reading Whitney and finite
  elements with hindsight},'' {\em IEEE Trans. Magn.}, vol.~43, pp.~1157--1160,
  2007.

\bibitem{Buffa11}
A.~Buffa, J.~Rivas, G.~Sangalli, and R.~Vazquez, ``Isogeometric discrete
  differential forms in three dimensions,'' {\em SIAM J. Numer. Anal.},
  vol.~49, pp.~818--844, 2011.

\bibitem{Back12}
A.~Back and E.~Sonnendr\"ucker, ``Spline discrete differential forms,'' {\em
  ESAIM: Proc.}, vol.~35, pp.~197--202, 2012.

\bibitem{albeverio90}
S.~Albeverio and B.~Zegarlinski, ``Construction of convergent simplicial
  approximations of quantum fields on riemannian manifolds,'' {\em Commun.
  Math. Phys.}, vol.~132, pp.~39--71, 1990.

\bibitem{albeverio95}
S.~Albeverio and J.~Schafer, ``Abelian chern-simons theory and linking numbers
  via oscillatory integrals,'' {\em J. Math. Phys.}, vol.~36, pp.~2157--2169,
  1995.

\bibitem{wilson07}
S.~O. Wilson, ``Cochain algebra on manifolds and convergence under
  refinement,'' {\em Topology Applicat.}, vol.~159, pp.~1898--1920, 2007.

\bibitem{wilson11}
S.~O. Wilson, ``Differential forms, fluids, and finite models,'' {\em Proc. Am.
  Math. Soc.}, vol.~139, pp.~2597--2604, 2011.

\bibitem{halvorsen12}
T.~G. Halvorsen and T.~M. Sorensen, ``Simplicial gauge theory and quantum gauge
  simulation,'' {\em Nucl. Phys. B}, vol.~854, pp.~166--183, 2012.

\bibitem{bossavitjapan}
A.~Bossavit, ``{Computational electromagnetism and geometry: (5) The "Galerkin
  Hodge"},'' {\em J. Jpn. Soc. Appl. Electromagn}, vol.~8, pp.~203--209, 2000.

\bibitem{Katz1998}
E.~Katz and U.~J. Wiese, ``{Lattice fluid dynamics from perfect discretizations
  of continuum flows},'' {\em Phys. Rev. E}, vol.~58, pp.~5796--5807, 1998.

\bibitem{MineP27a}
B.~He and F.~L. Teixeira, ``{A sparse and explicit FETD via approximate inverse
  Hodge (mass) matrix},'' {\em IEEE Microw. Wireless Comp. Lett.}, vol.~16,
  no.~6, pp.~348--350, 2006.

\bibitem{AdamsPRL97}
D.~H. Adams, ``{A doubled discretization of Abelian Chern-Simons theory},''
  {\em Phys. Rev. Lett.}, vol.~78, no.~22, pp.~4155--4158, 1997.

\bibitem{buffa}
A.~Buffa and S.~Christiansen, ``A dual finite element complex on the
  barycentric refinement,'' {\em Math. Comput.}, vol.~76, pp.~1743--1769, 2007.

\bibitem{GilletteCAD11}
A.~Gillette and C.~Bajaj, ``Dual formulations of mixed finite element methods
  with applications,'' {\em Comput. Aided Des.}, vol.~43, pp.~1213--1221, 2011.

\bibitem{Calcagni}
G.~Calcagni, D.~Oriti, and J.~Thurigen, ``Laplacians on discrete and quantum
  geometries,'' {\em Class. Quantum Grav.}, vol.~30, p.~125006, 2013.

\bibitem{Graf}
W.~Graf, ``Differential forms as spinors,'' {\em Annales de l'institut Henri
  Poincaré A: Physique théorique}, vol.~29, pp.~85--109, 1978.

\bibitem{AdamsPRD05}
D.~H. Adams, ``Fourth root prescription for dynamical staggered fermions,''
  {\em Phys. Rev. D}, vol.~72, p.~114512, 2005.

\bibitem{Friedan}
D.~Friedan, ``{A proof of the Nielsen-Ninomiya theorem},'' {\em Commun. Math.
  Phys.}, vol.~85, pp.~481--490, 1982.

\bibitem{Herbut}
I.~F. Herbut, ``{Time reversal, fermion doubling, and the masses of lattice
  Dirac fermions in three dimensions},'' {\em Phys. Rev. B}, vol.~83,
  p.~245445, 2011.

\bibitem{Raszillier}
H.~Raszillier, ``Lattice degeneracies for fermions,'' {\em J. Math. Phys.},
  vol.~25, pp.~1682--1693, 1984.

\bibitem{kanamori}
I.~Kanamori and N.~Kawamoto, ``{Dirac-K\"ahler femion with noncommutative
  differential forms on a lattice},'' {\em Nucl. Phys. B (Proc. Supl.)},
  vol.~129, pp.~877--879, 2004.

\bibitem{Susskind}
L.~Susskind, ``Lattice fermions,'' {\em Phys. Rev. D}, vol.~16, pp.~3031--3039,
  1977.

\bibitem{MineP1}
M.~G. do~Amaral, M.~Kischinhevsky, C.~A.~A. de~Carvalho, and F.~L. Teixeira,
  ``{An efficient method to calculate field theories with dynamical
  fermions},'' {\em Int. J. Mod. Phys. C}, vol.~2, no.~2, pp.~561--600, 1991.

\bibitem{Benn}
I.~M. Benn and R.~W. Tucker, ``{The Dirac equation in exterior form},'' {\em
  Comm. Math. Phys.}, vol.~98, pp.~53--63, 1985.

\bibitem{Beauce}
V.~de~Beauce, S.~Sen, and J.~C. Sexton, ``{Chiral Dirac fermions on the latice
  using Geometric Discretization},'' {\em Nucl. Phys. B (Proc. Supl.)},
  vol.~129, pp.~468--470, 2004.

\bibitem{AdamsPRL10}
D.~H. Adams, ``Theoretical foundation for the index theorem on the lattice with
  staggered fermions,'' {\em Phys. Rev. Lett.}, vol.~104, p.~141602, 2010.

\bibitem{Fillion}
F.~Fillion-Gourdeau and E.~L. A.~D. Bandrauk, ``{Numerical solution of the
  time-dependent Dirac equation in coordinate space without
  fermion-doubling},'' {\em Comp. Phys. Comm.}, vol.~183, pp.~1402--1415, 2012.

\bibitem{Berenger1994}
J.~P. Berenger, ``{A perfectly matched layer for the absorption of
  electromagnetic waves},'' {\em J. Comput. Phys.}, vol.~114, no.~2,
  pp.~185--200, 1994.

\bibitem{Chew_Weedon1994}
W.~C. Chew and W.~Weedon, ``{A 3D perfectly matched medium from modified
  Maxwell's equations with stretched coordinates},'' {\em Microwave Opt. Tech.
  Lett.}, vol.~7, no.~13, pp.~599--604, 1994.

\bibitem{TeixeiraMGWL1997a}
F.~L. Teixeira and W.~C. Chew, ``{PML-FDTD in cylindrical and spherical
  grids},'' {\em IEEE Microwave Guided Wave Lett.}, vol.~7, no.~9,
  pp.~285--287, 1997.

\bibitem{CollinoSIAM1998}
F.~Collino and P.~Monk, ``{The perfectly matched layer in curvilinear
  coordinates},'' {\em SIAM J. Sci. Computing}, vol.~19, pp.~2061--2090, 1998.

\bibitem{Sacks1995}
Z.~S. Sacks, D.~M. Kingsland, R.~Lee, and J.-F. Lee, ``{A perfectly matched
  anisotropic absorber for use as an absorbing boundary condition},'' {\em IEEE
  Trans. Antennas Propagat.}, vol.~43, no.~12, pp.~1460--1463, 1995.

\bibitem{TeixeiraMGWL1997b}
F.~L. Teixeira and W.~C. Chew, ``{Systematic derivation of anisotropic PML
  absorbing media in cylindrical and spherical coordinates},'' {\em IEEE
  Microwave Guided Wave Lett.}, vol.~7, no.~11, pp.~371--373, 1997.

\bibitem{TeixeiraMOTL1998}
F.~L. Teixeira and W.~C. Chew, ``{Analytical derivation of a conformal
  perfectly matched absorber for electromagnetic waves},'' {\em Microwave Opt.
  Technol. Lett.}, vol.~17, no.~4, pp.~231--236, 1998.

\bibitem{dondericiAP08}
B.~Donderici and F.~L. Teixeira, ``Conformal perfectly matched layer for the
  mixed finite-element time-domain method,'' {\em IEEE Trans. Antennas
  Propagat.}, vol.~56, pp.~1017--1026, 2008.

\bibitem{MineP18}
F.~L. Teixeira and W.~C. Chew, ``{On causality and dynamic stability of
  perfectly matched layers for FDTD simulations},'' {\em IEEE Trans. Microw.
  Theory Tech.}, vol.~47, no.~6, pp.~775--785, 1999.

\bibitem{MineP19}
F.~L. Teixeira and W.~C. Chew, ``{Complex space approach to perfectly matched
  layers: A review and some new developments},'' {\em Int. J. Num. Model.},
  vol.~13, no.~5, pp.~441--455, 2000.

\bibitem{Hockney}
R.~W. Hockney and J.~W. Eastwood, {\em Computer Simulation Using Particles}.
\newblock Bristol, U.K.: IOP Publishing, 1988.

\bibitem{esirkepov}
T.~Z. Ezirkepov, ``Exact charge conservation scheme for particle-in-cell
  simularion with an arbitrary form-factor,'' {\em Comp. Phys. Comm.},
  vol.~135, pp.~144--153, 2001.

\bibitem{ok06b}
Y.~Omelchenko and H.~Karimabadi, ``{Event-driven hybrid particle-in-cell
  simulation: A new paradigm for multi-scale plasma modeling},'' {\em J. Comp.
  Phys.}, vol.~216, pp.~153--178, 2006.

\bibitem{Mardahl}
P.~J. Mardahl and J.~P. Venboncoeur, ``{Charge conservation in electromagnetic
  PIC codes: Spectral comparison of Boris/DADI and Langdon-Marder methods},''
  {\em Comp. Phys. Comm.}, vol.~106, pp.~219--229, 1997.

\bibitem{Assous}
F.~Assous, ``A three-dimensional time-domain electromagnetic particle-in-cell
  code on unstructured grids,'' {\em Int. J. Model. Simul.}, vol.~29,
  pp.~279--284, 2009.

\bibitem{candel09}
A.~Candel, A.~Kabel, L.-Q. Lee, Z.~Li, C.~Ng, G.~Schussman, and K.~Ko, ``State
  of the art in electromagnetic modeling for the compact linear collider,''
  {\em J. Phys.: Conf. Ser.}, vol.~180, p.~012004, 2009.

\bibitem{squire12}
J.~Squire, H.~Qin, and W.~M. Tang, ``{Geometric integration of the
  Vlaslov-Maxwell system with a variational partcile-in-cell scheme},'' {\em
  Phys. Plasmas}, vol.~19, p.~084501, 2012.

\bibitem{Arnold2002}
D.~N. Arnold, ``{Differential complexes and numerical stability},'' in {\em
  Proceedings of the International Congress ofMathematicians, Beijing, Volume
  I: Plenary Lectures}, 2002.

\bibitem{munkres}
J.~R. Munkres, {\em Topology}.
\newblock Pearson, second~ed., 2000.

\bibitem{ChevalierAP97}
M.~W. Chevalier, R.~J. Luebbers, and V.~P. Cable, ``{FDTD local grid with
  material traverse},'' {\em IEEE Trans. Antennas Propagat.}, vol.~45,
  pp.~411--421, 1997.

\bibitem{WhiteMTT01}
M.~J. White, Z.~Yun, and M.~F. Iskander, ``{A new 3-D FDTD multigrid technique
  with dielectric traverse capabilities},'' {\em IEEE Trans. Microw. Theory
  Tech.}, vol.~49, no.~3, pp.~422--430, 2001.

\bibitem{christ12}
S.~H. Christiansen and T.~G. Halvorsen, ``{A simplicial gauge theory},'' {\em
  J. Math. Phys.}, vol.~53, p.~033501, 2012.

\bibitem{BossavitPIER32}
A.~Bossavit, ``{'Generalized finite differences' in computational
  electromagnetics},'' in {\em Geometric Methods for Computational
  Electromagnetics (PIER Series 32)} (F.~L. Teixeira, ed.), pp.~45--64,
  Cambridge, Mass.: EMW Publishing, 2001.

\bibitem{WeilandIJNM96}
P.~Thoma and T.~Weiland, ``{A consistent subgridding scheme for the finite
  difference time domain method},'' {\em Int. J. Num. Model.}, vol.~9,
  pp.~359--374, 1996.

\bibitem{RailtonEL97}
K.~M. Krishnaiah and C.~J. Railton, ``{Passive equivalent circuit of FDTD: An
  application of subgridding},'' {\em Electron. Lett.}, vol.~33, no.~15,
  pp.~1277--1278, 1997.

\bibitem{Schuhmann98}
R.~Schuhmann and T.~Weiland, ``Stability of the {FDTD} algorithm on
  nonorthogonal grids related to the spatial interpolation scheme,'' {\em IEEE
  Trans. Magn.}, vol.~34, no.~5, pp.~275--278, 1998.

\bibitem{chilton2008}
R.~A. Chilton, {\em H-, P- and T-Refinement Strategies for the
  Finite-Difference-Time-Domain (FDTD) Method Developed via Finite-Element (FE)
  Principles}.
\newblock PhD thesis, The Ohio State University, 2008.

\bibitem{YeeAP97}
K.~S. Yee and J.~S. Chen, ``{The finite-difference time-domain (FDTD) and the
  finite-volume time-domain (FVTD) methods in solving Maxwell's equations},''
  {\em IEEE Trans. Antennas Propagat.}, vol.~45, no.~3, pp.~354--363, 1997.

\bibitem{MattiussiPIER01}
C.~Mattiussi, ``{The geometry of time-stepping},'' in {\em Geometric Methods in
  Computational Electromagnetics, PIER 32} (F.~L. Teixeira, ed.), pp.~123--149,
  Cambridge, Mass.: EMW Publishing, 2001.

\bibitem{Weiland84}
T.~Weiland, ``{On the numerical solution of Maxwell's equations and
  applications in accelerator physics},'' {\em Particle Accelerators}, vol.~15,
  pp.~245--291, 1996.

\bibitem{Schuhmann00}
R.~Schuhmann and T.~Weiland, ``{Rigorous analysis of trapped modes in
  accelerating cavities},'' {\em Phys. Rev. Special Topics - Accelerators and
  Beams}, vol.~3, p.~122002, 2000.

\bibitem{codecasa04}
L.~Codecasa, V.~Minerva, and M.~Politi, ``Use of barycentric dual grids,'' {\em
  IEEE Trans. Magn.}, vol.~40, pp.~1414--1419, 2004.

\bibitem{schuhmann02}
R.~Schuhmann, P.~Schmidt, and T.~Weiland, ``{A new Whitney-based material
  operator for the finite-integration technique on triangular grids},'' {\em
  IEEE Trans. Magn.}, vol.~38, pp.~409--412, 2002.

\bibitem{bullo04}
M.~Bullo, F.~Dughiero, M.~Guarnieri, and E.~Tittonel, ``Isotropic and
  anisotropic electrostatic field computation by means of the cell method,''
  {\em IEEE Trans. Magn.}, vol.~40, pp.~1314--1317, 2004.

\bibitem{alotto06}
P.~Alotto, A.~D. Cian, and G.~Molinari, ``{A time-domain 3-D full-Maxwell
  solver based on the Cell Method},'' {\em IEEE Trans. Magn.}, vol.~42,
  pp.~799--802, 2006.

\bibitem{bullo06}
M.~Bullo, F.~Dughiero, M.~Guarnieri, and E.~Tittonel, ``Nonlinear coupled
  thermo-electromagnetic problems with the cell method,'' {\em IEEE Trans.
  Magn.}, vol.~42, pp.~991--994, 2006.

\bibitem{alotto08}
P.~Alotto, M.~Bullo, M.~Guarnieri, and F.~Moro, ``A coupled
  thermo-electromagnetic formulation based on the cell method,'' {\em IEEE
  Trans. Magn.}, vol.~44, pp.~702--705, 2008.

\bibitem{alotto10}
P.~Alotto, F.~Freschi, and M.~Repetto, ``{Multiphysics problems via the cell
  method: The role of Tonti diagrams},'' {\em IEEE Trans. Magn.}, vol.~46,
  pp.~2959--2962, 2010.

\bibitem{codecasa}
L.~Codecasa, R.~Specogna, and F.~Trevisan, ``Discrete geometric formulation of
  admittance boundary conditions for frequency domain problems over tetrahedral
  dual grids,'' {\em IEEE Trans. Antennas Propagat.}, vol.~60, pp.~3998--4002,
  2012.

\bibitem{SteinbergJCP95}
M.~Shashkov and S.~Steinberg, ``{Support operator finite difference algorithms
  for general elliptic problems},'' {\em J. Comp. Phys.}, vol.~118,
  pp.~131--151, 1995.

\bibitem{ShashkovJCP99}
J.~M. Hyman and M.~Shashkov, ``{Mimetic discretizations for Maxwell's
  equations},'' {\em J. Comp. Phys.}, vol.~151, pp.~881--901, 1999.

\bibitem{ShashkovSIAM99}
J.~M. Hyman and M.~Shashkov, ``{The orthogonal decompostion theorems for
  mimetic finite difference mehods},'' {\em SIAM J. Num. Analys.}, vol.~36,
  pp.~788--818, 1999.

\bibitem{ShashkovANM97}
J.~M. Hyman and M.~Shashkov, ``{Adjoint operators for the natural
  discretizations of the divergence, gradient, and curl in logically
  rectangular grids},'' {\em Appl. Num. Math.}, vol.~25, pp.~413--442, 1997.

\bibitem{ShashkovPIER01}
J.~M. Hyman and M.~Shashkov, ``{Mimetic finite difference methods for Maxwell's
  equations and the equations of magnetic diffusion},'' in {\em Geometric
  Methods in Computational Electromagnetics, PIER 32} (F.~L. Teixeira, ed.),
  pp.~89--121, Cambridge, Mass.: EMW Publishing, 2001.

\bibitem{CastilloANM02}
J.~Castillo and T.~McGuinness, ``{Steady-state diffusion problems on
  non-trivial domain: Support operator method integrated with direct optimized
  grid generation},'' {\em Appl. Num. Math.}, vol.~40, pp.~207--218, 2002.

\bibitem{lipnikov06}
K.~Lipnikov, M.~Shashkov, and D.~Svyatskiy, ``The mimetic finite difference
  discretization of diffusion problem on unstructured polyhedral meshes,'' {\em
  J. Comp. Phys.}, vol.~211, pp.~473--491, 2006.

\bibitem{brezzi10}
F.~Brezzi and A.~Buffa, ``Innovative mimetic discretizations for
  electromagnetic problems,'' {\em J. Comp. Appl. Math.}, vol.~234,
  pp.~1980--1987, 2010.

\bibitem{robidoux11}
N.~Robidoux and S.~Steinberg, ``A discrete vector calculus in tensor grids,''
  {\em Comp. Meth. Appl. Math.}, vol.~1, pp.~1--44, 2011.

\bibitem{lipnikov11}
K.~Lipnikov, M.~Manzini, F.~Brezzi, and A.~Buffa, ``The mimetic finite
  difference method for the 3d magnetostatic field problems on polyhedral
  meshes.,'' {\em J. Comp. Phys.}, vol.~230, pp.~305--328, 2011.

\bibitem{Lipnikov}
K.~Lipnikov, G.~Manzini, and M.~Shashkov, ``Mimetic finite difference method,''
  {\em J. Comp. Phys.}, 2013 (to appear).

\bibitem{arnold02}
D.~N. Arnold, ``Differential complexes and numerical stability,'' in {\em
  Proceedings of the International Congress of Mathematicians}, vol.~I: Plenary
  Lectures, (Beijing, China), 2002.

\bibitem{arnold06a}
D.~N. Arnold, P.~B. Bochev, R.~B. Lehoucq, R.~A. Nicolaides, and M.~Shashkov,
  eds., {\em Compatible Spatial Discretizations}.
\newblock IMA Volumes in Mathematics and its Applications, Springer-Verlag,
  2006.

\bibitem{white06}
D.~White, J.~Koning, and R.~Rieben, ``{Development and application of
  compatible discretizations of Maxwell's equations},'' in {\em Compatible
  Discretization of Partial Differential Equations}, Springer-Verlag, 2006.

\bibitem{bochev06}
P.~Bochev and M.~Gunzburger, ``Compatible discretizations of second-order
  elliptic problems,'' {\em J. Math. Sci.}, vol.~136, pp.~3691--3705, 2006.

\bibitem{boffi07}
D.~Boffi, ``Approximation of eigenvalues in mixed form, discrete compactness
  property, and application to hp mixed finite elements,'' {\em Comp. Meth.
  Appl. Mech. Eng.}, vol.~196, pp.~3672--3681, 2007.

\bibitem{Bochev12}
P.~Bochev, H.~C. Edwards, R.~C. Kirby, K.~Peterson, and D.~Ridzal, ``{Solving
  PDEs with Intrepid},'' {\em Sci. Programming}, vol.~20, pp.~151--180, 2012.

\bibitem{nedelec80}
J.~C. Nedelec, ``{Mixed finite elements in $R^3$},'' {\em Numer. Math.},
  vol.~35, pp.~315--341, 1980.

\bibitem{hiptmairMC99}
R.~Hiptmair, ``Canonical construction of finite elements,'' {\em Math. Comp.},
  vol.~68, pp.~1325--1346, 1999.

\bibitem{Guil}
V.~W. Guillemin and S.~Sternberg, {\em Supersymmetry and Equivariant de Rham
  Theory}.
\newblock Berlin: Springer, 1999.

\bibitem{arnold06b}
D.~N. Arnold, R.~S. Falk, and R.~Winther, ``Finite element exterior calculus,
  homological techniques, and applications,'' {\em Acta Numerica}, vol.~15,
  pp.~1--155, 2006.

\bibitem{Yavari08}
A.~Yavari, ``On geometric discretization of elasticity,'' {\em J. Math. Phys.},
  vol.~49, p.~022901, 2008.

\bibitem{Bossavitchap}
A.~Bossavit, ``{Mixed finite elements and the complex of Whitney forms},'' in
  {\em The Mathematics of Finite Elements and Applications} (J.~R. Whiteman,
  ed.), pp.~137--144, Academic Press, 1988.

\bibitem{wong95}
M.-F. Wong, O.~Picon, and V.~F. Hanna, ``{A finite element method based on
  Whitney forms to solve Maxwell equations in the time domain},'' {\em IEEE
  Trans. Magn.}, vol.~31, pp.~1618--1621, 1995.

\bibitem{feliziani98}
M.~Feliziani and F.~Maradei, ``{Mixed finite-difference/Whitney-elements
  time-domain (FD/WE-TD) method},'' {\em IEEE Trans. Magn.}, vol.~34,
  pp.~3222--3227, 1998.

\bibitem{castillo04}
P.~Castillo, J.~Koning, R.~Rieben, and D.~White, ``A discrete differential
  forms framework for computational electromagnetics,'' {\em Comp. Meth. Eng.
  Sci.}, vol.~5, pp.~331--346, 2004.

\bibitem{rieben05}
R.~N. Rieben, G.~H. Rodrigue, and D.~A. White, ``{A higher order mixed vector
  finite element method for solving the time dependent Maxwell equations on
  unstructured grids},'' {\em J. Comp. Phys.}, vol.~204, pp.~490--519, 2005.

\bibitem{Desbrun03}
M.~Dsebrun, A.~N. Hirani, and J.~E. Mardsen, ``Discrete exterior calculus for
  variational problem in computer vision and graphics,'' in {\em Proc. 42nd
  IEEE Conf. Decision Control}, (Maui, Hawaii, USA), pp.~4902--4907, 2003.

\bibitem{Hirani03}
A.~N. Hirani, {\em Discrete Exterior Calculus}.
\newblock PhD thesis, Calif. Inst. Technol., 2003.

\bibitem{Desbrun05}
M.~Desbrun, A.~N. Hirani, M.~Leok, and J.~E. Mardsen, ``Discrete exterior
  calculus.'' available from arXiv.org/math.DG/0508341, 2005.

\bibitem{Gillette09}
A.~Gillette, ``Notes on discrete exterior calculus,'' tech. rep., Univ. Texas
  at Austin, 2009.

\bibitem{perot}
J.~B. Perot, ``Discrete conservation properties of unstructures mesh schemes,''
  {\em Annual Rev. Fluid Mech.}, vol.~2011, pp.~299--318, 2011.

\bibitem{Kotiuga08}
P.~R. Kotiuga, ``Theoretical limitation of discrete exterior calculus in the
  context of computational electromagnetics,'' {\em IEEE Trans. Magn.},
  vol.~44, pp.~1162--1165, 2008.

\end{thebibliography}

\end{document}